\documentclass[11pt,a4paper,oneside]{article}
\usepackage{jheppub}
\pdfoutput=1
\usepackage{myterms}
\usepackage{mathrsfs}
\usepackage[english]{babel}
\usepackage{amsmath,amssymb,amsfonts, bm,bbm,slashed}
\usepackage{graphicx}
\usepackage[sort&compress]{natbib}
\usepackage{xcolor}
\usepackage[normalem]{ulem}
\usepackage{hyperref}
\usepackage{cleveref}
\definecolor{red}{rgb}{1.0, 0, 0}
\usepackage{enumerate}
\usepackage{epsfig, subfigure}
\usepackage{setspace}
\usepackage{booktabs, tabularx}
\usepackage{units}
\usepackage[utf8]{inputenc}
\usepackage{lineno}
\usepackage{feynmp-auto}               
\usepackage{placeins}
\usepackage{slashed}
\usepackage{appendix}

\hypersetup{
	pdfnewwindow=true,   
	colorlinks=true,     
	linkcolor=blue,      
	citecolor=blue,      
	filecolor=blue,      
	urlcolor=blue        
}

\allowdisplaybreaks

\bibpunct{[}{]}{,}{n}{}{,}

\newcommand{\vev}[1]{{\langle #1 \rangle}}

\newcommand{\GeV}{\,\text{GeV}}
\newcommand{\TeV}{\,\text{TeV}}

\newcommand{\sigeff}{\ensuremath{\sigma_{\textrm{eff}}} }
\newcommand{\geff}{\ensuremath{g_{\textrm{eff}}} }

\makeatletter
\gdef\@fpheader{}
\makeatother

\begin{document}
	

\title{Collider Signatures of Coannihilating Dark Matter in Light of the B-Physics Anomalies}	

\author[a]{Michael~J.~Baker,}    
\emailAdd{michael.baker@unimelb.edu.au}
\author[b]{Darius A. Faroughy,}
\emailAdd{faroughy@physik.uzh.ch}
\author[c]{Sokratis Trifinopoulos}
\emailAdd{sokratis.trifinopoulos@ts.infn.it}

\affiliation[a]{ARC Centre of Excellence for Dark Matter Particle Physics, School of Physics, The University of Melbourne, Victoria 3010, Australia}

\affiliation[b]{Physik-Institut, Universit\"at Z\"urich, 8057 Z\"urich, Switzerland}

\affiliation[c]{INFN, Sezione di Trieste, SISSA, Via Bonomea 265, 34136, Trieste, Italy}

\date{\today}
\preprint{ZU-TH-43/21}

	
\abstract{
Motivated by UV explanations of the $B$-physics anomalies, we study a dark sector containing a Majorana dark matter candidate and a coloured coannihilation partner, connected to the Standard Model predominantly via a $U_1$ vector leptoquark.  A TeV~scale $U_1$ leptoquark, which couples mostly to third generation fermions, is the only successful single-mediator description of the $B$-physics anomalies. After calculating the dark matter relic surface, we focus on the most promising experimental avenue: LHC searches for the coloured coannihilation partner. We find that the coloured partner hadronizes and forms meson-like bound states leading to resonant signatures at colliders reminiscent of the quarkonia decay modes in the Standard Model. By recasting existing dilepton and monojet searches we exclude coannihilation partner masses less than 280 GeV and 400 GeV, respectively. Since other existing collider searches do not significantly probe the parameter space, we propose a new dedicated search strategy for pair production of the coloured partner decaying into $bb\tau\tau$ final states and dark matter particles. This search is expected to probe the model up to dark matter masses around 600 GeV with current luminosity. 
}

\maketitle

\section{Introduction}
\label{sec:introduction}

The Standard Model (SM) of particle physics is the most accurate description of the fundamental particles and their interactions. There are, however, solid experimental and theoretical reasons to postulate the existence of New Physics (NP), e.g., astrophysical observations have established the presence of dark matter in the universe~\cite{Bertone:2004pz}. Since dark matter cannot be accounted for by the particle content of the SM, its nature remains one of the biggest mysteries in modern physics.

At the high-energy frontier, no definite Beyond the SM (BSM) signals have emerged from the full set of run-II LHC data. However, in the last decade a large number of low-energy flavour experiments performed at $B$-factories and LHCb have uncovered indirect hints of lepton flavour universality violation in $B$-mesons decays in $b\to s\ell^+\ell^-$ and $b\to c\ell^\pm\nu$ transitions. Interestingly, these deviations from the SM, known as the $B$-physics anomalies, seem to point towards a new boson at the TeV~scale, which couples predominantly to third generation fermions. The only models that can simultaneously accommodate all $B$-physics anomalies while satisfying the rich low-energy phenomenology and high-$p_T$ constraints feature leptoquarks as the  mediators of the NP effects~\cite{Hiller:2014yaa,Gripaios:2014tna,Alonso:2015sja,Fajfer:2015ycq,Calibbi:2015kma,Bauer:2015knc,Barbieri:2015yvd,Faroughy:2016osc,Becirevic:2016yqi,Bhattacharya:2016mcc,Assad:2017iib,Calibbi:2017qbu,Blanke:2018sro,Becirevic:2018afm,Kumar:2018kmr}.

In particular, the vector leptoquark $U_1$ with SM quantum numbers $({\bf 3},{\bf 1},2/3)$ has emerged as the only single-mediator solution~\cite{Buttazzo:2017ixm,Angelescu:2018tyl,Angelescu:2021lln}. Within the scope of ultraviolet (UV) complete frameworks, this vector leptoquark can arise as a gauge boson of a spontaneously broken gauge group containing $SU(3)_c$ colour as a subgroup. The minimal phenomenologically viable group is $SU(4)\times SU(3)' \times SU(2)_L\times U(1)_{T_R^3}$; models based on this group are called `4321 models'~\cite{DiLuzio:2017vat,Bordone:2017bld,Greljo:2018tuh,Bordone:2018nbg,DiLuzio:2018zxy,Cornella:2019hct,Fuentes-Martin:2020bnh,King:2021jeo}. These particular UV models predict other heavy gauge bosons, e.g., a $Z^\prime$ and a colour octet (coloron), that give rise to a rich set of collider signatures that can currently be tested at the LHC~\cite{Baker:2019sli,Cornella:2021sby}.  In addition, these models introduce a wide range of new fermionic states, which may also lead to new collider signatures.  

Inspired by these UV constructions, we explore the possibility that dark matter is contained within a new fermionic multiplet, which would be intimately connected to QCD through a larger gauge symmetry.\footnote{For other works that propose a connection between the $B$-physics anomalies and dark matter, see refs.~\cite{Sierra:2015fma, Belanger:2015nma, Kawamura:2017ecz, Ko:2017quv, Fuyuto:2017sys, Cline:2017qqu, Azatov:2018kzb, Choi:2018stw, Singirala:2018mio, Hati:2018fzc, Falkowski:2018dsl, Baek:2018aru, Hutauruk:2019crc,Trifinopoulos:2019lyo,Guadagnoli:2020tlx,Carvunis:2020exc,Huang:2020ris,DEramo:2020sqv,Arcadi:2021glq,Becker:2021sfd,Arcadi:2021cwg}.}  In particular, we investigate whether dark matter could be a neutral remnant of a fermionic multiplet of the broken gauge symmetry, much in the same way leptons are colourless remnants of a multiplet unifying quarks and leptons (as in the Pati-Salam model~\cite{Pati:1974yy}). This would suggest that there may be other dark sector fermionic partners similar in mass to the dark matter particle that carry colour and electric charge.  The $U_1$ leptoquark and other new gauge bosons may then mediate interactions between the dark sector and the SM. 

Furthermore, a Majorana mass term generated by higher-dimensional operators induces a pseudo-Dirac scenario~\cite{DeSimone:2010tf} where the would-be Dirac dark matter fermion splits into two Majorana fermions, the lighter being a dark matter candidate. We show that the quasi-degeneracy of the dark matter candidate and its coloured partner can be naturally preserved within our general UV framework, and can have a significant impact on the thermal freeze-out of dark matter and collider phenomenology. For example, the relic abundance will be determined by coannihilating effects in the early universe. The coloured coannihilating partner also hadronizes and forms QCD bound states, which may lead to unexpected signatures at the LHC.
 
The best probe of the dark sector is LHC production of two coannihilation partners.  These can either form a bound state with each other, which then decays into dileptons via electroweak interactions, or each partner can decay via a leptoquark, leading to a $b\bar b\tau^-\tau^+$ and missing transverse energy signature. Since the $b$ and $\tau$ particles will be soft, due to the quasi-degeneracy of the dark sector, existing ATLAS~\cite{ATLAS:2018uni,ATLAS:2019qpq} and CMS~\cite{CMS:2018txo,CMS:2018iye} searches are not sensitive. We propose a dedicated search in this channel which exploits a novel observable: the ratio between the visible and missing energy of the process. 
By simulating signal and background events, we estimate the reach at the LHC (with run-II luminosity) of both a cut-based and a multi-variate analysis, finding a significant improvement when using a boosted decision tree (BDT) classifier.

The paper is structured as follows. In \cref{sec:model} we first motivate the simplified model from a UV perspective, and then present the Lagrangian relevant at the TeV scale. The coannihilation effects are described in \cref{sec:relic-surface} and the parameters required to reproduce the observed dark matter relic abundance are calculated. In \cref{sec:psi_pheno} we describe the hadronization of the coannihilating partner and the resulting meson spectroscopy. We then examine the main experimental constraints set by existing collider searches and derive bounds on the relevant parameter space. Finally, in \cref{sec:new_search}, we present our dedicated search strategy for the pair production of the coannihilating partner.

\section{Theoretical Motivation and the Simplified Model}
\label{sec:model}

Although we will study a simplified model, we first motivate Majorana dark matter coannihilating with a slightly heavier colour triplet partner from a UV perspective. Under the SM group $SU(3)_c \times SU(2)_L \times U(1)_Y$ the dark sector states will have the representations
\begin{equation}\label{eq:reps}
  \chi\sim({\bf 1,1},0)\,,\qquad\qquad  \psi\sim({\bf 3,1},2/3) \,,
\end{equation}
and couple directly to the vector leptoquark $U_1\sim(\mathbf{3}, \mathbf{1}, 2/3)$.  
There are two main ways of accounting for a multi-TeV vector leptoquark: 
gauge models and strongly interacting models~\cite{Baker:2019sli}.  In this work we will assume a gauge model explanation, 
where the leptoquark is a massive gauge boson associated with the spontaneous breaking of a gauge symmetry, $G_\text{NP} \supset G_\text{SM}$.  \par

\subsection{General Setup}
\label{sec:generalsetup}

We assume that in addition to the SM matter content, there is also a heavy vector-like fermion transforming under some representation of $G_\text{NP}$ that decomposes as
\begin{align}
\label{eq:vlf}
\mathcal{X} &=\, \chi\oplus\psi\oplus \ldots
\end{align}
under the SM gauge group. Although the simplest example is to take $\mathcal{X} =\, (\psi^1 \, \psi^2 \, \psi^3 \, \chi)^T$ transforming as the fundamental of $SU(4)$, we emphasise that the setup is more general and \cref{eq:vlf} is strictly all that is required. After the breaking of $G_\text{NP}$, $\psi$ becomes coloured under $SU(3)_c$ and $\chi$ remains colourless and electrically neutral, with quantum numbers given by \cref{eq:reps}. We imagine that $\mathcal{X}$ is charged under some stabilising symmetry while the SM particles are not, so the components of $\mathcal{X}$ constitute a dark sector and the neutral component leads to a dark matter candidate. The breaking of $G_{\rm NP}\to G_{\rm SM}$ around the TeV scale typically requires a set of scalar fields uncharged under the stabilising symmetry that we denote collectively as $\Omega$. Since $\mathcal{X}$ is a Dirac field by construction, we introduce non-renormalizable interactions so that the neutral component $\chi$ gives rise to Majorana dark matter.\footnote{Although the field content in our setup is similar to that considered in ref.~\cite{Guadagnoli:2020tlx}, the dark matter candidate in ref.~\cite{Guadagnoli:2020tlx} is a Dirac fermion. For Dirac dark matter, the dominant interaction that connects the dark and SM sectors is via a $Z'$ gauge boson and the strongest experimental limits come from direct detection experiments.  Once appropriate non-renormalizable operators are included, the dark matter candidate becomes a Majorana fermion, the dominant interaction is mediated by a $U_1$ vector leptoquark and direct detection constraints are negligible. Instead, collider searches are the best probes of the motivated parameter space.}  
This then points towards a higher scale (beyond the TeV) associated with fermion number violation. 

Before the spontaneous symmetry breaking of $G_\text{NP}$, the effective Lagrangian for the dark sector up to dimension $d\!=\!5$ is
\begin{align}
\label{eq:LDS}
\mathcal{L}^\text{DS}_\text{eff} &=\, i \overline{\mathcal{X}} \slashed{D} \mathcal{X} - m_\mathcal{X} \overline{\mathcal{X}}\mathcal{X} - \sum_n\frac{c^{(5)}_n}{\Lambda}\mathcal{O}^{(5)}_n\,,
\end{align}
where $\Lambda$ is a large cut-off scale associated with fermion number violation and $\mathcal{O}^{(5)}_n$ are effective operators with Wilson coefficients $c^{(5)}_n$. Since $\mathcal{X}$ is a vector-like fermion, the operators $\mathcal{O}^{(5)}_n$ fall into two possible categories: (i) operators of the form $\overline{\mathcal{X}} F \mathcal{X}$ that conserve fermion number, and (ii) Weinberg type operators of the form $\overline{\mathcal{X}^c}F^\prime \mathcal{X}$ that violate fermion number by two units. Here $F$ and $F^\prime$ are $d\!=\!2$ scalar operators that are bilinears in the fields $\Omega$ responsible for spontaneous symmetry breaking. Once these scalar fields develop vacuum expectation values (vevs) the $\mathcal{O}^{(5)}_n$ terms in \cref{eq:LDS} will lead to both Dirac and Majorana masses for the components of $\mathcal{X}$ of the form $\vev{F}/\Lambda$ and $\vev{F^\prime}/\Lambda$, respectively. These masses will be suppressed because the scale $\Lambda$ associated with fermion number violation is taken to be much larger than the $G_\text{NP}$ breaking scale. In fact, we will assume that $\Lambda^2\gg \vev{F},\vev{F^\prime},m^2_\mathcal{X}$. In \cref{sec:4321DM} we provide a simple 4321 extension that gives rise to these two types of $d\!=\!5$ effective operators. 

Note that the Lagrangian \cref{eq:LDS} is not completely general since there could also be $d\!=\!4$ Yukawa terms of the form $\overline{\mathcal{X}}\Omega\mathcal{X}$, if the scalar sector $\Omega$ contains states with the correct quantum numbers (e.g., the adjoint scalar $\Omega_{15}$ in the 4321 model discussed in \cref{app:4321}). Once these fields acquire vevs, $\overline{\mathcal{X}}\Omega\mathcal{X}$ could lead to a mass splitting of order $\vev{\Omega}$ between the $\psi$ and $\chi$ components of the $\mathcal{X}$ multiplet, while still preserving colour. We here assume that these type of $d\!=\!4$ terms are not present or are negligible (e.g., due to small Yukawa couplings) and that the only relevant sources of mass-splitting at tree level between the $\mathcal{X}$ components arise from $d=5$ operators or higher. Higher order effective operators with $d\ge6$ can also be included in \cref{eq:LDS}, but these will have little impact on the dark sector phenomenology and for this reason can be dropped.

After the breaking of $G_\text{NP}$, the scenario described above will lead to the following mass terms for the dark sector particles: 
\begin{align}
\label{eq:DSmass}
\mathcal{L}^\text{DS}_\text{mass} = &\, -m_\psi\overline{\psi} \psi - m_\chi \overline{\chi} \chi - \frac{1}{2}\left( m_L\, \overline{\chi_L^c} \chi_L + m_R\, \overline{\chi_R^c} \chi_R + \text{h.c.} \right)\,,
\end{align} 
where $m_\psi$ and $m_\chi$ are two Dirac masses roughly of the same order, satisfying $m_\psi=m_\mathcal{X}$ and $|m_\psi-m_\chi|\sim \vev{F} / \Lambda \ll m_\mathcal{X}$, and $m_{L}$ and $m_R$ are two small Majorana masses of order $ \sim \vev{F^\prime} / \Lambda$, such that $ m_{L,R}\ll m_\mathcal{X}$. Note that in general we expect $m_{L,R}^2 \lesssim \vev{F^\prime}$ as the Majorana mass terms in \cref{eq:DSmass} violate the $G_{NP}$ symmetry, so they may only break it softly at scales lower than the breaking scale. The resulting mass eigenstates $\chi_{1,2}$ are two pseudo-Dirac fermions \cite{Tucker-Smith:2001myb,DeSimone:2010tf} with quasi-degenerate masses $m_{\chi_{1,2}}$, given by
\begin{align}\label{eq:chi1}
\chi_1 \simeq\, \frac{i}{\sqrt{2}} \left( \chi - \chi^c \right)\,, &\ \ \ \ \ m_{\chi_1}\simeq m_\chi-\frac{m_L+m_R}{2}\,, \\ \label{eq:chi2}
\chi_2\simeq\, \frac{1}{\sqrt{2}} \left( \chi + \chi^c \right) \,,&\ \ \ \ \  m_{\chi_2}\simeq m_\chi+\frac{m_L+m_R}{2} \,,
\end{align}
up to leading order in $|m_L-m_R|/m_\chi$ and the Majorana condition $\chi_{1,2}\simeq\chi_{1,2}^c$ is satisfied up to the same order. In the mass basis, the Lagrangian reads
\begin{align}\label{eq:LDS2}
    \mathcal{L}^\text{DS} \ =\ \bar\psi(i\slashed{D}-m_{\psi})\psi\ +\, \sum_{i=1,2}\, \frac{1}{2}\bar\chi_i(i\slashed{D}-m_{\chi_i})\chi_i\,.
\end{align}
At tree level, the spectrum of the dark sector satisfies $m_{\chi_1}\simeq m_{\chi_2}\simeq m_\psi$ with mass splittings of order $\vev{F}/\Lambda\sim\vev{F^\prime}/\Lambda$. In what follows we will be interested in the compressed spectrum scenario $m_{\chi_1} \lesssim m_\psi \lesssim m_{\chi_2}$, so that the lightest stable state $\chi_1$ is a dark matter candidate.   Small deviations from \cref{eq:chi1,eq:chi2} can occur at loop level if additional heavy states couple differently to $\psi$ and $\chi$. For example, the heavy gauge bosons in the 4321 models produce order 10\% mass splittings between $\psi$ and $\chi$~\cite{Guadagnoli:2020tlx}.  

In summary, we see that relatively general considerations lead to models with a dark sector containing a coloured partner $\psi$ and pseudo-Dirac pairs $\chi_1$ and $\chi_2$, where  
$\chi_1$, $\chi_2$ and $\psi$ have a compressed spectrum.

\subsection{UV Realisation: Majorana Dark Matter in a 4321 Model}
\label{sec:4321DM}

In order to further motivate the general setup presented above, we present a concrete example of a UV model which gives rise to the dark sector effective Lagrangian \cref{eq:LDS}. For this, we extend the matter field content of the standard (or flavoured) 4321 model (see \cref{app:4321} for more details) with a dark sector containing a $\mathbb{Z}_2$--odd vector-like fermion $\mathcal{X}\sim({\bf4,1,1},+1/2)$ with components  $\mathcal{X}=(\psi,\chi)^T$, and a $\mathbb{Z}_2$--odd right-handed fermion singlet $S_R\sim({\bf1, 1,1},0)$. The role of this exact $\mathbb{Z}_2$ symmetry is to stabilise the dark matter. The $\mathbb{Z}_2$--even SM fields are exactly as in the 4321 models, as shown in \cref{tab:4321}. The scalar sector of 4321 contains the $\mathbb{Z}_2$--even state $\Omega_{1}\sim({\bf \bar4,1,1},-1/2)$, which gives rise to Yukawa interactions between the dark sector fields. The Lagrangian of the dark sector reads
\begin{equation}\label{eq:Lag4321DS0}
    \mathcal{L}^{\rm DS} =\overline{\mathcal{X}}(i\slashed{D}-m_\mathcal{X})\mathcal{X}+i\overline{S_R}\slashed{\partial}S_R-\left(\frac{M_S}{2}\overline{S_R^c}S_R +\lambda\,\overline{S_R^c}\Omega_1^T\mathcal{X}_R+\lambda^\prime\,\overline{\mathcal{X}}_L\Omega_1^* S_R+\hc \right)\,.
\end{equation}
It is convenient to rotate the singlet fields into the Majorana basis and work with the field $N$ defined by
\begin{align}
    N\equiv e^{i\frac{\theta}{2}}S_R+e^{-i\frac{\theta}{2}}S_R^c\,,
\end{align}
such that $N\!=\!N^c$ and where the $\theta$ phase corresponds to the argument of the Majorana mass, $M_S\!=\!|M_S|\, \text{exp}(i\theta)$. We also define the couplings $\lambda_R\!\equiv\!\lambda\, \text{exp}(-i \theta/2)$ and $\lambda_L\!\equiv\!\lambda^\prime\, \text{exp}(-i \theta/2)$. We can now rewrite the terms involving $S_R$ in \cref{eq:Lag4321DS0} as
\begin{align}
\mathcal{L}^{\rm DS} \supset \frac{1}{2}\overline{N}(i\slashed{\partial}-M_N)N &
-\frac{1}{2}\overline{N}\Big[\Omega_1^T(\lambda_R\mathcal{X}_R+\lambda_L^*\mathcal{X}_L)+ \Omega_1^\dagger(\lambda_R^*\mathcal{X}_R^c+\lambda_L\mathcal{X}_L^c)\Big] \notag\\
&
-\frac{1}{2}\Big[(\lambda_R\overline{\mathcal{X}^c_R}+\lambda_L^*\overline{\mathcal{X}^c_L})\Omega_1+(\lambda_R^*\overline{\mathcal{X}}_R+\lambda_L\overline{\mathcal{X}}_L)\Omega_1^*\Big]N\,.
\end{align}
The Majorana mass $M_N\!\equiv\!|M_S|$ is assumed to be much larger than the Dirac mass of $\mathcal{X}$, $M_N\gg m_\mathcal{X}$, since it is generated at a very high scale where fermion number is violated. This allows us to integrate out the $N$ field, giving rise to the effective Lagrangian given in \cref{eq:LDS} with the following $d=5$ operators and corresponding Wilson coefficients:
\begin{align}
\mathcal{O}^{(5)}_{RR}
&=\overline{\mathcal{X}^c_R}\,\Omega_1\Omega_1^T\mathcal{X}_R\,, 
&\frac{c^{(5)}_{RR}}{\Lambda}
&=-\frac{\lambda_R^2}{2M_N}
\,,\\
\mathcal{O}^{(5)}_{LL}
&=\overline{\mathcal{X}_L}\,\Omega_1^*\Omega_1^\dagger\mathcal{X}^c_L\,,
&\frac{c^{(5)}_{LL}}{\Lambda}
&=-\frac{\lambda_L^2}{2M_N}
\,,\\
\mathcal{O}^{(5)}_{LR}
&=\overline{\mathcal{X}}_L\,\Omega_1^*\Omega_1^T\mathcal{X}_R\,,
&\frac{c^{(5)}_{LR}}{\Lambda}
&=-\frac{\lambda_L\lambda_R}{M_N}\,,
\end{align}
and their Hermitian conjugates.  The first two operators $\mathcal{O}^{(5)}_{RR,LL}$ are Weinberg type operators and violate fermion number by two units, leading to Majorana mass terms after spontaneous symmetry breaking, while $\mathcal{O}^{(5)}_{LR}$ conserves fermion number, leading to a Dirac mass term. After $\Omega_1$ obtains a vev, $\langle \Omega_1 \rangle = (0,0,0,\omega_1/\sqrt{2})^T$, the 4321 group breaks down to the SM gauge group and the physical masses in \cref{eq:LDS2} are given by
\begin{align}
m_\psi&=m_\mathcal{X}\,,\label{eq:mpsi}\\
m_{\chi_1}&\simeq m_\mathcal{X}-\frac{(\lambda_L - \lambda_R)^2\,\omega_1^2}{4 M_N}\,,\label{eq:mchi1}\\
m_{\chi_2}&\simeq m_\mathcal{X}+\frac{(\lambda_L + \lambda_R)^2\,\omega_1^2}{4 M_N}\,.\label{eq:mchi2}
\end{align}
Notice that for $\lambda_L\ne\lambda_R$ this UV model reproduces the desired compressed mass spectrum with the correct ordering $m_{\chi_1} < m_\psi < m_{\chi_2}$.  
Furthermore, the eigenstates $\chi_{1,2}$ naturally arise as a pair of pseudo-Dirac fermions since the $\chi_1$--$\chi_2$ mass splitting terms $m_{L,R} = \lambda_{L,R}^2\omega_1^2/2M_N$ are suppressed by the seesaw mechanism in the limit $M_N\gg \omega_1$.

\subsection{The Simplified Model}
\label{sec:simplified-model}

In our analysis, we will study a simplified model motivated by the above scenario.  We will parametrise the mass splittings as
\begin{align}
\Delta_\psi &\equiv\, \frac{m_\psi - m_{\chi_1}}{m_{\chi_1}} \,, \qquad \qquad
\Delta_{\chi_2} \equiv\, \frac{ m_{\chi_2} - m_{\chi_1}}{m_{\chi_1}} \,,
\end{align}
focusing on the regime where $0 \leq \Delta_\psi \lesssim 0.3$, as explained in the next section, and where $\Delta_{\chi_2} \gtrsim 2 \Delta_\psi$ (motivated by \cref{eq:mpsi,eq:mchi1,eq:mchi2}).

The covariant derivative terms in $\mathcal{L}^\text{DS}$, \cref{eq:LDS2}, contain couplings to gauge bosons which become massive after $G_\text{NP}$ breaking.  These will act as mediators between the dark sector and the SM. As discussed in the introduction, the $B$--anomalies suggest the existence of a $U_1$ vector leptoquark. Once this is included, closure of the algebra means that there must also be a 
heavy $Z'$~\cite{Baker:2019sli}. Although there could be further gauge bosons, such as a coloron, we assume that these do not significantly impact the dark sector phenomenology.
The complete set of new fields we introduce in our simplified model is shown in \cref{tab:particles}.  
We will see that the relic abundance and LHC phenomenology is driven by 
$\chi_1$, $\psi$ and $U_1$. In the notation of ref.~\cite{Baker:2015qna} which classifies co-annihilation models and their LHC signatures, this is model ST2 
 (note that \cite{Baker:2015qna} uses a different hyper-charge convention 
 to the one we employ here). 
The interaction terms of the new gauge bosons are
\begin{align}
\mathcal{L}^{\rm int}_{U} = &\, 
\frac{g_U}{\sqrt{2}}\, U_1^{\mu,\alpha} [
\beta_L^{ij}\,\overline{q_L^{i,\alpha}} \gamma_\mu \ell_L^j + 
\beta_R^{ij}\,\overline{ d_R^{i,\alpha}} \gamma_\mu e_R^j + 
\beta_{D_1}\,\overline{ \chi_1} \gamma_\mu \psi^\alpha + 
\beta_{D_2}\,\overline{ \chi_2} \gamma_\mu \psi^\alpha
] +{\rm h.c.}\,, \label{eq:LU}
\\
\mathcal{L}^{\rm int}_{Z^\prime} =&\, 
\frac{g_{Z^\prime}}{2\sqrt{6}}\,Z^{\prime\,\mu} [
\zeta_q^{ij}\,\overline{ q^i_L} \gamma_\mu q_L^j + 
\zeta_u^{ij}\,\overline{ u^i_R} \gamma_\mu u_R^j  + 
\zeta_d^{ij}\,\overline{ d^i_R} \gamma_\mu d_R^j - 
3\,\zeta_\ell^{ij}\,\overline{ \ell^i_L} \gamma_\mu \ell_L^j  -
3\,\zeta_e^{ij}\,\overline{ e_R^i} \gamma_\mu e_R^j \,+ \notag \\
&\, 
\zeta_\psi \overline{\psi} \gamma_\mu \psi +
i \zeta_\chi \overline{\chi}_2 \gamma_\mu \chi_1
]\,,
 \label{eq:Lint}
 \end{align}
where the numerical prefactors are motivated by the 4321 model~\cite{DiLuzio:2018zxy}.
Note that the Majorana fields $\chi_1$ and $\chi_2$ satisfy $\overline{\chi_i}\gamma_\mu \chi_j = - \overline{\chi_j}\gamma_\mu \chi_i$ for $i,j\in\{1,2\}$, so the vector currents $\overline{\chi_i}\gamma_\mu \chi_i$ vanish and the above Lagrangian is Hermitian.

\begin{table}
  \begin{center}
    \begin{tabular}{@{\hspace{1em}}c@{\hspace{2em}}c@{\hspace{2em}}c@{\hspace{2em}}c@{\hspace{1em}}}
    \hline\hline
      Field  &      Type     & SM QN & $\mathbb{Z}_2$ \\
      \hline
      $\chi_1$ & Majorana &  $(\mathbf{1},\mathbf{1},0)$ &    $-1$   \\
      $\chi_2$ & Majorana &  $(\mathbf{1},\mathbf{1},0)$ &    $-1$     \\
      $\psi$ &  Dirac &  $(\mathbf{3},\mathbf{1},2/3)$ &   $-1$     \\
      \hline
      $Z'$    & Gauge &   $(\mathbf{1},\mathbf{1},0)$ &   $+1$    \\
      $U_1$    & Gauge &   $(\mathbf{3},\mathbf{1},2/3)$ &   $+1$    \\
    \hline\hline
    \end{tabular} 
  \caption{The new fermions and bosons we introduce in our simplified model with their quantum numbers under the SM gauge group and the stabilising $\mathbb{Z}_2$ symmetry.
    }
  \label{tab:particles}
  \end{center}
\end{table}

The $\mathcal{O}(1)$ coupling $g_U$ and the masses $m_U$, $m_{\chi_1}$, $m_\psi$ are the parameters which dominate the determination of the dark matter abundance and collider physics signatures we consider. 
For these parameters, the key relation that provides a solution to the $B$-physics anomalies is~\cite{Cornella:2021sby}
 \begin{equation}
 \label{eq:gUmU}
g_U  = (1.1 \pm 0.2) \times\left(  \frac{m_U}{2\TeV} \right) \,.
\end{equation} 

For the $\beta$ and $\zeta$ couplings we work in the flavour basis where the $SU(2)_L$ SM fermion doublets are aligned with the down-quark sector,
\begin{align}
q_L^i=
\begin{pmatrix}
V^*_{ji}\, u_L^j\\
d_L^i
\end{pmatrix}\,,
\qquad\qquad
\ell_L^i=
\begin{pmatrix}
\nu^i\\
e_L^i
\end{pmatrix}\,,
\end{align}
where $V$ is the CKM matrix.
In this basis, motivated by the $B$--anomalies and minimality, we assume the following structure for the $\beta$ couplings
\begin{equation}
\label{eq:minCoup}
\begin{aligned}
\beta_{L} = 
\left(\begin{array}{ccc} 
0 & 0 & \beta_L^{d\tau} \\ 
0 & \beta_{L}^{s \mu} &  \beta_{L}^{s \tau}\\ 
0 & \beta_{L}^{b \mu} &  1
\end{array}\right)
\,,\qquad\qquad 
\beta_{R} = 
\left(\begin{array}{ccc} 
0 & 0 & 0 \\ 
0 & 0 & 0\\ 
0 & 0 &  -1
\end{array}\right)
\,, \qquad\qquad
\beta_{D_1} = \beta_{D_2} = 1 \,.
\end{aligned}
\end{equation}
Non-zero values for $\beta_L^{ij}$ are needed to address the $B$--anomalies, and may arise when connecting this simplified model to UV models addressing the structure of the SM Yukawa couplings~\cite{Cornella:2019hct,King:2021jeo}. 
While these parameters only have a marginal influence on the dark matter abundance and the collider physics signatures we consider in this work, for definiteness we take $\beta_{L}^{s \tau } =  -  \beta_{L}^{b \mu}  \approx 0.11$ and 
 $\beta_{L}^{s \mu} = - \beta_{L}^{d \tau} \approx  0.02$~\cite{Cornella:2019hct}.

For the $Z'$ couplings, we take $\zeta_q^{33} = \zeta_u^{tt} = \zeta_d^{bb} = \zeta_\ell^{33} = \zeta_e^{\tau\tau} = \zeta_\psi = \zeta_\chi = 1$ and all others equal to zero.  Although we will take $g_{Z'} = g_U$ and $m_{Z'} = m_U/\sqrt{2}$, these parameters do not have a significant influence on the dark matter abundance or LHC signatures considered (as long as $g_{Z'}$ isn't too large and the $Z'$ isn't too light, as detailed below).
Finally, in addition to the kinetic term for the leptoquark we also include non-minimal interactions between the leptoquark and the SM gauge fields appearing in gauge models,
\begin{align}
\mathcal{L}_{U}&\supset\, 
-\frac{1}{2} U_{1\,\mu\nu}^\dagger U_1^{\mu\nu}
-ig_s U_{1\,\mu}^\dagger T^a U_{1\,\nu} G^{a\,\mu\nu}
-ig_Y \frac{2}{3} U_{1\,\mu}^\dagger U_{1\,\nu}\,B^{\mu\nu} \,,
\end{align}
where $T^a=\lambda^a/2$ and $\lambda^a$ ($a=1,\dots,8)$ are the Gell-Mann matrices.

\section{Dark Matter Relic Surface}
\label{sec:relic-surface}

We now turn to the dark matter relic abundance.  
The presence of the coannihilating partner, $\psi$, can heavily impact the relic abundance 
of the dark matter candidate, $\chi_1$.  
We will restrict the parameter space by insisting that the model 
produces the observed relic abundance via thermal freeze-out. 

First, to clearly describe the physics of coannihilation, we will imagine that the masses of the dark sector (DS) particles $\chi_1$, $\chi_2$ and $\psi$ are above the electroweak scale 
and that the leptoquark and $Z'$ are heavy enough not to be present during $\chi_1$ freeze-out 
(we allow them to be lighter in our numerical work below). 
If $m_\psi, m_{\chi_2} \gg m_{\chi_1}$ then all processes connecting $\chi_1$ to the thermal 
bath are heavily suppressed.  As such, $\chi_1$ will freeze-out when relativistic and will 
have a relic abundance many orders of magnitude too large to match observations (it will overclose the universe). 
However, the situation is very different if $m_{\chi_1} \sim m_\psi \sim m_{\chi_2}$. 
In this case the abundances of $\chi_1$, $\chi_2$ and $\psi$ are 
similarly Boltzmann suppressed during freeze-out of $\chi_1$.  
The conversion processes $\text{DS SM} \to \text{DS SM}$ 
have one rare, heavy particle and one bath particle in the initial state,  
so they have a rate exponentially larger than dark sector 
annihilation processes $\text{DS DS} \to \text{SM SM}$, which requires two rare particles.  
In our model $\overline{\chi}_1 \chi_1 \to \text{SM SM}$ is absent while both $\overline{\chi}_1 \psi \to \text{SM SM}$ (via an $s$-channel leptoquark) and 
$\overline{\psi} \psi \to \text{SM SM}$ (via QCD processes) are efficient.  
This means that although $\chi_1$ cannot efficiently annihilate with $\overline{\chi}_1$, it can annihilate with $\overline{\psi}$ or be efficiently depleted by first converting into $\psi$, via $\chi_1 \text{ SM} \to \psi \text{ SM}$, which then annihilates via $\overline{\psi} \psi \to \text{SM SM}$.  
This coannihilation effect will be relevant if the mass splitting between $\psi$ and $\chi_1$ is small~\cite{Griest:1990kh},
\begin{align}
\Delta_\psi \lesssim 0.3\,,  
\end{align}
and becomes more important as $\Delta_\psi$ shrinks.  
Similarly, if $\Delta_{\chi_2} \lesssim 0.3$ then $\overline{\chi_1} \chi_2 \to \text{SM SM}$ via an $s$-channel $Z'$ may be efficient.  
However, since $\Delta_{\chi_2} \gtrsim 2 \Delta_\psi$, $g_{Z'} \sim g_U$ and $m_{Z'} \sim m_U$ and since the relic abundance is predominantly determined by the most efficient process (which is $\overline{\chi}_1 \psi \to \text{SM SM}$ or $\overline{\psi} \psi \to \text{SM SM}$ in the parameter space of interest), $\chi_2$ does not play an important role in setting the relic abundance.  We do however keep it in our numerical work.

To accurately calculate the relic abundance of $\chi_1$, the Boltzmann equation must track the abundances of $\chi_1$, $\chi_2$ and $\psi$.  This coupled differential equation can be written as a single differential equation~\cite{Griest:1990kh}, which is equivalent to the usual Boltzmann equation for a single species but with the annihilation cross-section replaced by
\begin{align} \label{eq:sigeff}
 \sigeff &= \sum_{ij} \frac{g_i g_j}{\geff^2} \sigma_{ij}
(1+\Delta_{i})^\frac{3}{2}(1+\Delta_{j})^\frac{3}{2}
e^{-x(\Delta_i+\Delta_j)} \, ,
\end{align}
where $i, j \in \{\chi_1, \chi_2, \psi\}$ index the dark sector particles (note that $\Delta_{\chi_1} = 0$), 
$g_i$ is the number of degrees of freedom of particle $i$ ($g_{\chi_1} = g_{\chi_2} = 2$, $g_\psi = 4$ in our model),
the cross-section is $\sigma_{ij} = \sigma(i j \rightarrow B B')$ where $B$ and $B'$ are particles in the thermal bath 
and $x = m_{\chi_1} / T$. The effective number of degrees of freedom is given by 
\begin{align}
 \geff &= \sum_{i} g_i(1+\Delta_i)^\frac{3}{2} e^{-x\Delta_i} \, .
\end{align}
We see that, for instance, when $ \sigma_{{\chi_1} \psi}\gg \sigma_{\overline{\chi}_1 \chi_1}$ and 
both $\Delta_\psi$ and $g_\text{eff}$ are not too large, then 
\begin{align}
\sigma_\text{eff} \gg  \sigma_{\overline{\chi}_1 \chi_1}\,.
\end{align} 
Since $\Omega_{\chi_1} h^2 \sim 1/ \langle \sigma v \rangle$, coannihilation effects then reduce the relic abundance of $\chi_1$.

\begin{figure}
  \centering
  \includegraphics[width=0.7\textwidth]{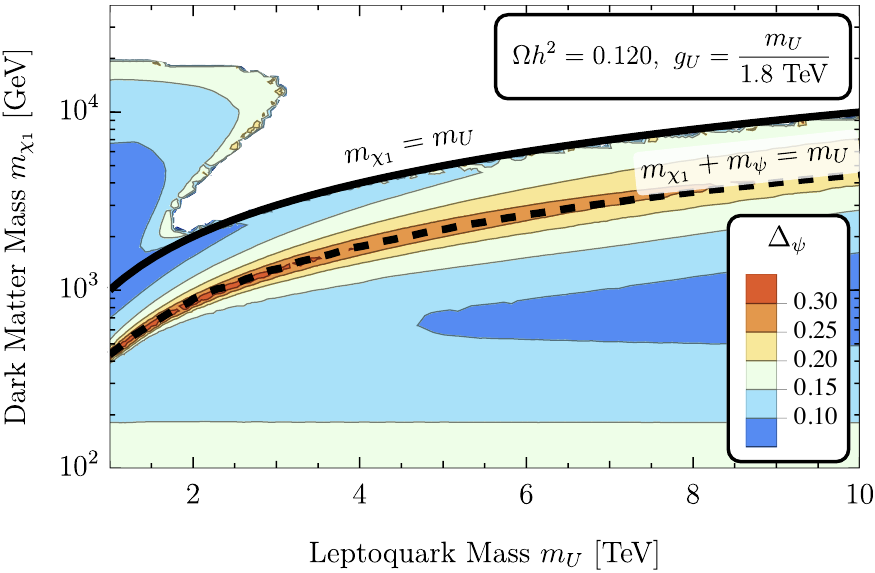}
  \caption{The mass splitting $\Delta_\psi$ that yields the observed 
  dark matter relic abundance as a function of the leptoquark mass $m_U$ and the dark matter 
  mass $m_{\chi_1}$. 
  While we set $\Delta_{\chi_2} = 2 \Delta_\psi$, $g_{Z'} = g_U$ and $m_{Z'} = m_U / \sqrt{2}$ in our numerics, 
the result is only weakly dependent on these parameters (as long as $\overline{\chi}_1 \chi_2 \to \text{SM SM}$ 
via an $s$-channel $Z'$ is subdominant).
  Below the solid black line $m_{\chi_1} < m_U$, while the dotted black line shows 
  the resonant region where $m_{\chi_1} + m_\psi = m_U$.  
  The observed dark matter relic abundance cannot be obtained in 
  the white region.
  }
  \label{fig:relic-surface}
\end{figure}

To calculate the relic abundances we use model files written with
\texttt{FeynRules\,v2.3} \cite{Alloul:2013bka} and 
solve the Boltzmann equations using \texttt{micrOMEGAs\,v5}~\cite{Belanger:2018mqt}.  
In principle, Sommerfeld corrections and the effects of bound state formation 
lead to corrections to the perturbative cross-sections of coloured particle annihilation 
in the early universe.  
However, for the case of fermionic triplets in the TeV mass range, these corrections are 
negligible~\cite{Buschmann:2016hkc,Keung:2017kot,ElHedri:2017nny} and we 
do not include them.  
We use the relation between $g_U$ and $m_U$ given in \cref{eq:gUmU} 
and determine the value of the mass splitting $\Delta_\psi$ which will result in the 
observed relic abundance~\cite{Aghanim:2018eyx},
\begin{align}
\Omega_\text{DM} h^2 = 0.120 \pm 0.001 \, .
\end{align}
The required value of $\Delta_\psi$ is shown in \cref{fig:relic-surface}.  
While we set $\Delta_{\chi_2} = 2 \Delta_\psi$, $g_{Z'} = g_U$ and $m_{Z'} = m_U / \sqrt{2}$, 
we have checked that 
the result is only weakly dependent on these parameters.
We see that the observed relic abundance can be obtained over a large 
mass range of $\chi_1$ ($100\GeV \lesssim m_{\chi_1} \lesssim 20\TeV$) for mass splittings 
in the range $0.05 \lesssim \Delta_\psi \lesssim 0.35$. 
We do not consider leptoquark masses below $1.5\TeV$ due to collider constraints, 
or above $10\TeV$, since $g_U$ becomes non-perturbative (due to \cref{eq:gUmU}).

We first discuss the region with $m_{\chi_1} < m_U$ (below the solid black line).  
Along the resonant line $m_{\chi_1} + m_\psi = m_U$ the cross-section for the 
$s$-channel process $\chi_1 \psi \to \text{SM SM}$ is large and so a large mass 
splitting is required to achieve the observed relic abundance.  
For a fixed $m_U$, as $m_{\chi_1}$ reduces the process goes further off resonance, so a smaller mass splitting is required to compensate (to keep $\sigma_\text{eff}$ approximately constant).  
This continues until $\psi \psi \to \text{SM SM}$ becomes the dominant process at small $m_{\chi_1}$.  Once this happens the required mass splitting grows as $m_{\chi_1}$, and so $m_\psi$, reduces.  In the region $m_U - m_\psi < m_{\chi_1} < m_U$, the mass splitting becomes smaller further from the resonance line, as the process again goes further off resonance.

When $m_{\chi_1} > m_U$ (above the solid black line), in much of the parameter space the dominant process 
is $\chi_1 \chi_1 \to U_1 U_1$ via a $t$-channel $\psi$ 
(when $m_U \ll m_{\chi_1}$, the leptoquark is still abundant as $\chi_1$ freezes-out, so this process can efficiently annihilate $\chi_1$).  
In this case the effective cross-section, \cref{eq:sigeff}, is not exponentially sensitive to $\Delta_\psi$, so small variations in 
$m_\psi$ cannot yield the observed relic abundance. 
This is indicated by the white region.  
Only when $m_U \lesssim 3\TeV$ do the processes $\chi_1 \psi \to B B'$ and $\psi \psi \to B B'$, where $B$ and $B'$ are bath particles,  
compete with $\chi_1 \chi_1 \to U_1 U_1$, so that $\Delta_\psi$ can be adjusted to obtain the observed relic abundance.
Above $m_{\chi_1} \sim 20\TeV$ the observed relic abundance cannot be obtained for any $\Delta_\psi$ or $m_U$.

\section{Coloured Coannihilation Partner Phenomenology}
\label{sec:psi_pheno}

Since the dark matter candidate in our model is a Majorana fermion, the direct and indirect detection constraints are negligible~\cite{Baker:2015qna,Buschmann:2016hkc,Baker:2018uox}.  As such, the strongest constraints on the simplified model parameter space will come from collider searches for the new particles. Collider searches for the leptoquark and $Z'$ have been well studied, and constrain the leptoquark to being heavier than $\sim1.7\TeV$~\cite{Baker:2019sli,Cornella:2021sby}. Although the presence of the dark sector may weaken these limits, the leptoquark and $Z'$ will still have significant branching ratios to SM particles.  The lightest dark sector particle, the dark matter candidate $\chi_1$, is a gauge singlet and only couples to the SM via the heavy leptoquark and $Z'$, so searches for the dark matter candidate directly are challenging. However, as discussed in \cref{sec:model}, the coannihilation partner $\psi$ is similar in mass to the dark matter candidate and is a colour triplet, making it directly accessible at hadron colliders. For this reason we focus on searches for $\psi$ production. 

After we fix the couplings as discussed in \cref{sec:simplified-model}, the model parameters relevant for collider searches are the masses of the dark matter candidate, $m_{\chi_1}$, the coannihilation partner, $m_\psi$, and the leptoquark, $m_U$.  In \cref{sec:relic-surface} we found the $\psi$ mass which leads to the observed dark matter relic abundance through thermal freeze-out, as shown in \cref{fig:relic-surface}.  In our collider analysis we use this result to eliminate $m_\psi$, allowing us to present our results in the $m_{\chi_1}$--$m_U$ plane.

Since $\psi$ is an unstable coloured fermion, it is important to compare its width with the QCD scale to see if it is likely to hadronize. We are primarily interested in the case where the $\mathbb{Z}_2$--odd fermions are significantly heavier than the SM particles, but lighter than the $U_1$ leptoquark (since the LHC will not be sensitive to dark sector particles heavier than a few TeV). Due to $\mathbb{Z}_2$--parity conservation, the dominant $\psi$ decay channels are the three-body processes $\psi \to \chi_1 b \tau$ and $\psi \to \chi_1 t \nu_\tau$, which are mediated by an off-shell leptoquark.   Due to the compressed spectrum, $m_\psi \sim m_{\chi_1}$, the width of $\psi$ is expected to be small. A straightforward computation (see \cref{eq:decay_width}) yields the partial width
\begin{align}\label{eq:psiwidth}
\Gamma_{\psi \to \chi_1 b \tau} \lesssim10^{-6}\, \text{GeV} \,\left(\frac{m_\psi}{\mathrm{TeV}}\right)^5 \,,
\end{align}
for the coannihilation regime $\Delta_\psi<0.3$ and for generic leptoquark couplings satisfying low energy constraints.\footnote{We obtain an expression similar to \cref{eq:psiwidth} for the partial width into the $\chi_1 t\nu_\tau$ channel.} These suppressed widths for a TeV scale $\psi$ suggest that it has enough time to hadronize into QCD bound states before it decays.

\subsection{$\Psi$--Meson Spectroscopy}

\begin{figure}[t]
\centering
\includegraphics[width=0.7\textwidth]{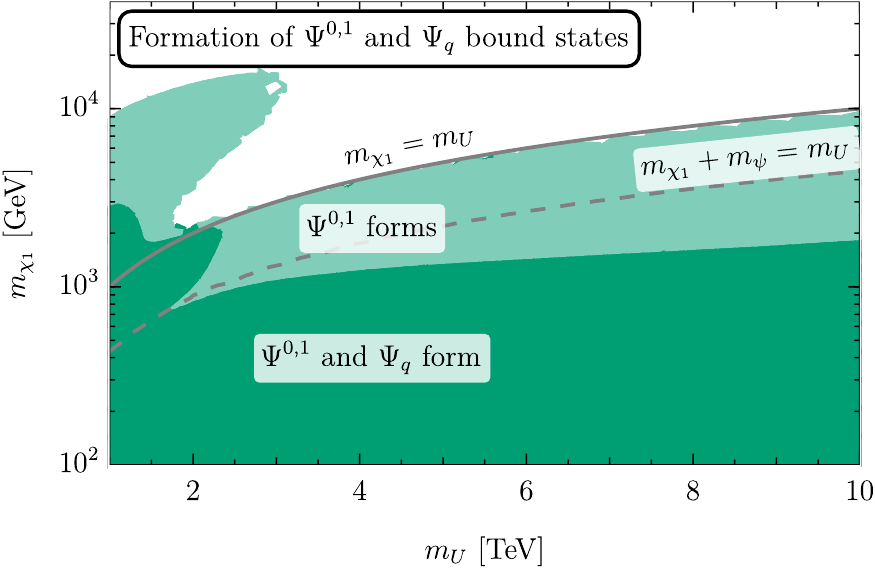}
\caption{Regions of the $m_{\chi_1} - m_{U}$ plane where $\psi$ hadronizes, with $\Delta_\psi$ fixed to yield the observed dark matter relic abundance. In the light green region the lifetime of $\psi$ is long enough for `psionium' states $(\psi\bar{\psi})$ to form, while in the dark green region formation of `open-psi' mesons $(\psi\bar{q})$ also occurs.
}
\label{fig:hadronization}
\end{figure}

There are two types of mesonic bound states: `psionium' states $(\psi\bar{\psi})$ and `open-psi' states $(\psi\bar{q})$, where $q$ is a light SM quark. Their formation can be described, at leading order, by non-relativistic QCD using a modified coulomb potential similar to the hydrogen model~\cite{Fabiano:1997xh}, see \cref{app:hadronization} for details. In this framework, the criteria for bound state formation is that the time it takes for the fermion pair to complete one rotation, $t_R$, (defined in \cref{eq:tR}) must be larger than their intrinsic lifetimes, in our case $\tau_\psi = \Gamma_{\psi}^{-1}$. In \cref{fig:hadronization} we show the regions of parameter space, on the relic surface, where the $\psi$ lifetime is sufficiently long for the psionium states $(\psi\bar{\psi})$ (denoted $\Psi^{0,1}$, see below) and open-psi states $(\psi\bar{q})$  (denoted $\Psi_q$) to form. Note that the lifetimes of these bound states are not long enough for them to be considered stable on collider scales.

The coannihilating partner $\psi$ thus gives rise to a rich spectrum of QCD bound states above the electroweak scale. The spectrum is expected to have a pattern similar to the bound state mesons containing charm or bottom quarks in the SM. 
These new heavy bound states can be classified using spin, parity, charge-conjugation and the new $\mathbb{Z}_2$--parity. We use the usual notation $J^{PC}_\pm$ with an additional $\pm$ subscript to indicate its $\mathbb{Z}_2$--parity. All psionia states are $\mathbb{Z}_2$--even, and the lightest psionium states are the $S$-wave pseudoscalar $ 0^{-+}_+$ followed by the vector $1^{--}_+$ (in analogy to the $\eta_c$ and $J/ \psi$ mesons in the charm sector of the SM, respectively). We name the ground state pseudoscalar state $\Psi^0$ and the lightest vector state $\Psi^1$. The open-psi bound states $(\psi\bar q)$, which we denote as $\Psi_q$, are all $\mathbb{Z}_2$--odd resonances analogous to the $D_q$--mesons in the SM.

\subsection{Collider Signatures}
\label{sec:coll_signatures}

 Given that $\psi$ carries both colour and charge, it can be produced in large numbers at hadron colliders, mainly via QCD interactions. Single $\psi$ production is forbidden by $\mathbb{Z}_2$--parity invariance, making $pp\to \psi\bar\psi$ the main production mechanism at the LHC. At threshold, the pair of $\psi$ particles will bind to form $\Psi$--mesons, leading to either single production of $\Psi^{0,1}$ states, or pair production of $\Psi_q$ states. In \cref{tab:modes} we show the main production and decay mechanisms for $\Psi^{0,1}$ and $\Psi_q$ at the LHC.  Note that there is no QCD $q\bar q\to \Psi^{0,1}$ production or $\Psi^{0,1} \to q\bar q$ decay due to angular momentum conservation and since $\Psi^1$ is a colour singlet~\cite{Kats:2012ym}.

\begin{table}[t]
  \begin{center}
    \begin{tabular}{@{\hspace{1em}}ccc@{\hspace{1em}}}
    \hline\hline
    Type & Production & Decay \\
    \hline
         & $gg\to \Psi^0$                       & $\Psi^0\to gg$        \\
     QCD & $gg\to \Psi^1 g$                     & $\Psi^1\to ggg$       \\
         & $q\bar q\,, gg\to \Psi_q\bar \Psi_q$ & $\Psi^1\to \gamma gg$ \\
    \hline
        &                         & $\Psi^0\to \gamma\gamma$  \\
     EW & $q\bar q\to \Psi^{0,1}$ & $\Psi^1\to q\bar q$       \\
        &                         & $\Psi^1\to \ell^+ \ell^-$ \\
    \hline
     NP & $bg\to \chi_1 \Psi_q\,\tau^\pm$ & $\Psi_q\to \chi_1 b\,\tau^\pm$ \\
    \hline\hline
    \end{tabular}
\caption{The main $\Psi$-meson production and decay modes for each type of interaction at the LHC.}
\label{tab:modes}
\end{center}
\end{table}

\subsubsection{Search for Psionium in the Dilepton Channel}
\label{sec:psionium-dilepton-bounds}

\begin{figure}
  \centering
	 \includegraphics[width=0.4\textwidth]{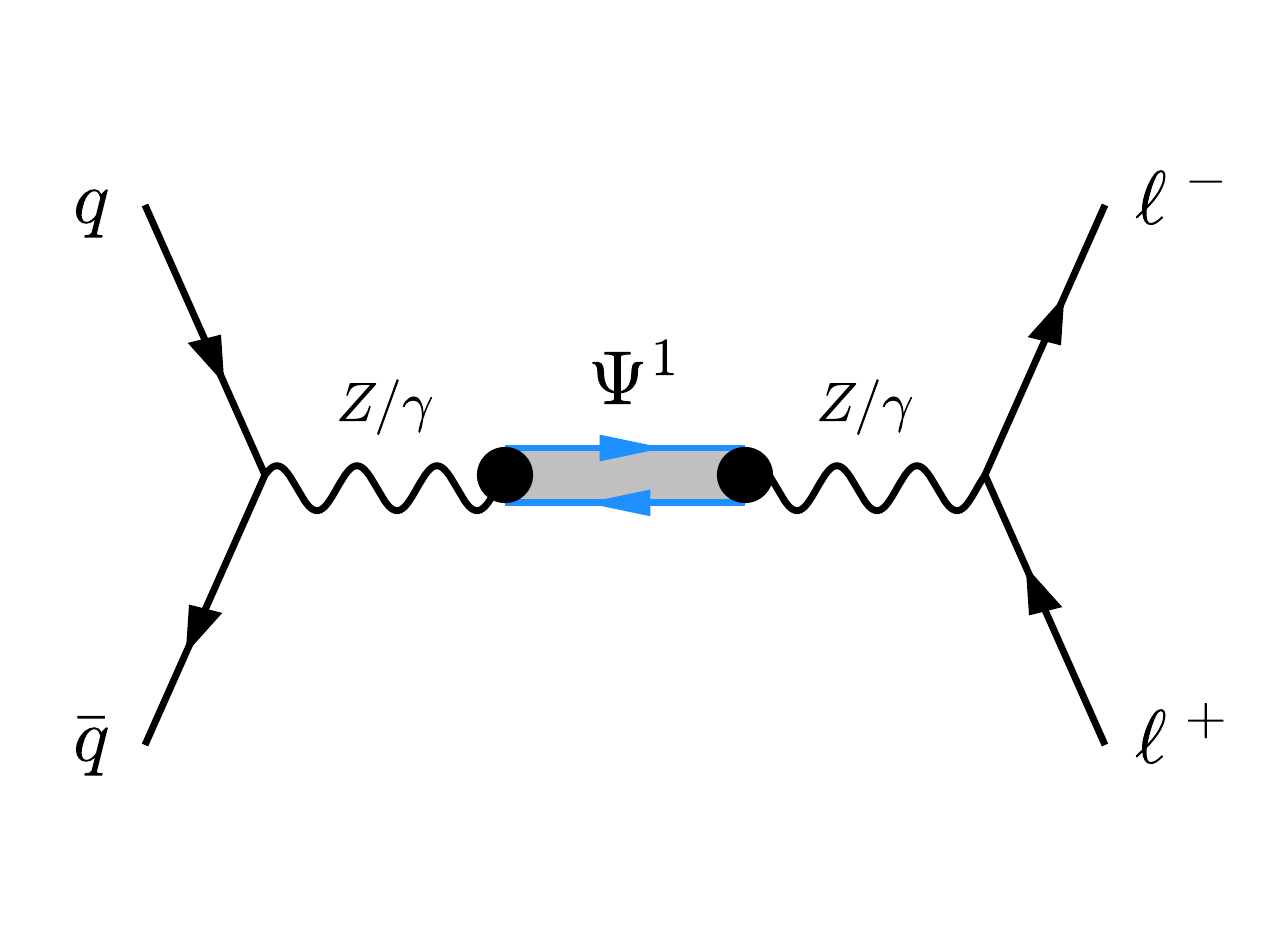} \quad
	 \includegraphics[width=0.4\textwidth]{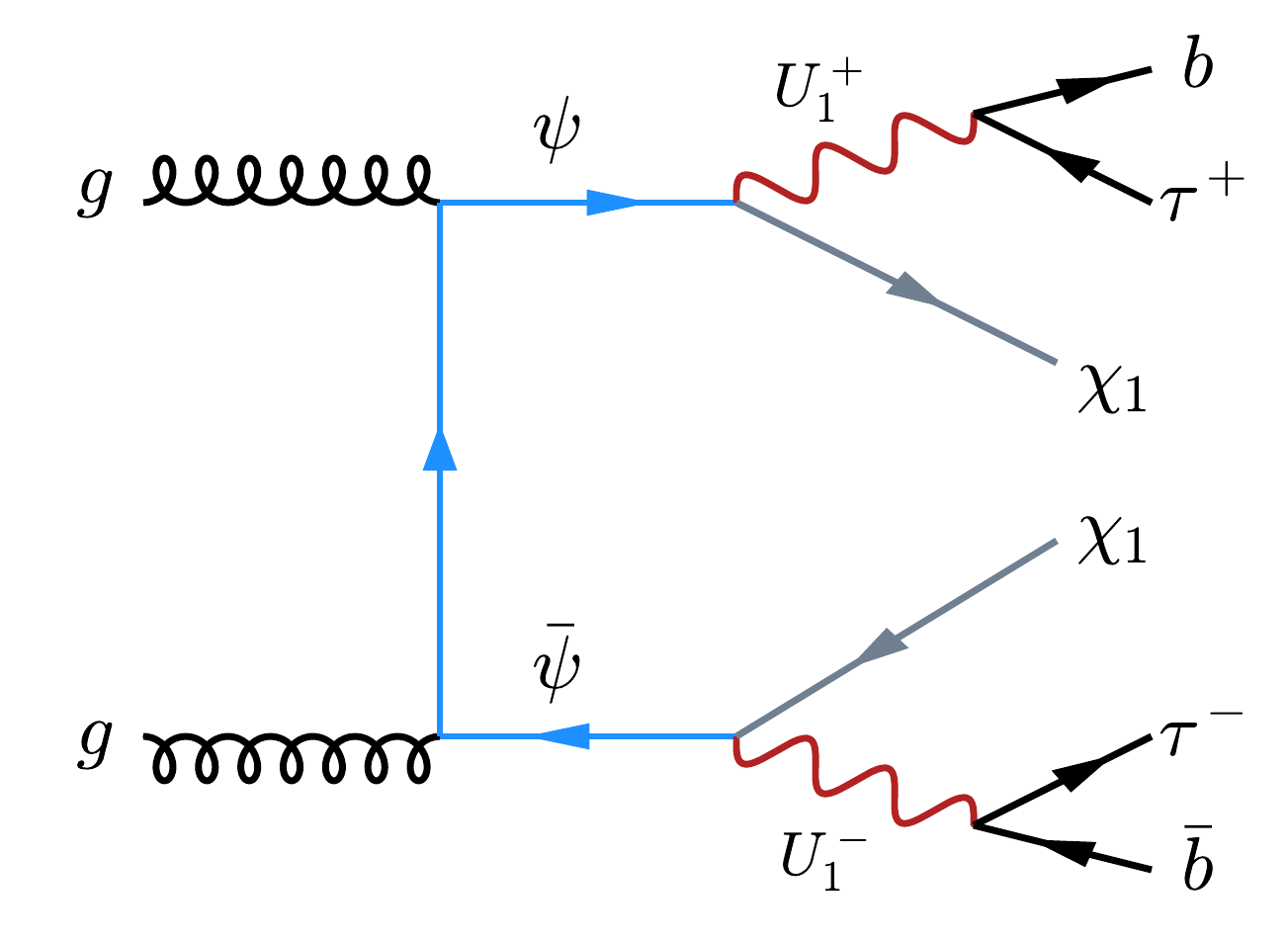}
  \caption{(Left) Electroweak production of $\Psi^1$ and its subsequent decay into two leptons. (Right) A representative diagram for pair production of $\psi$, which both decay into $b\tau\chi_1$.}
  \label{fig:psionium_production_decay}
\end{figure}

\begin{figure}
  \centering
  \begin{tabular}{cc}
  \includegraphics[width=0.6\textwidth]{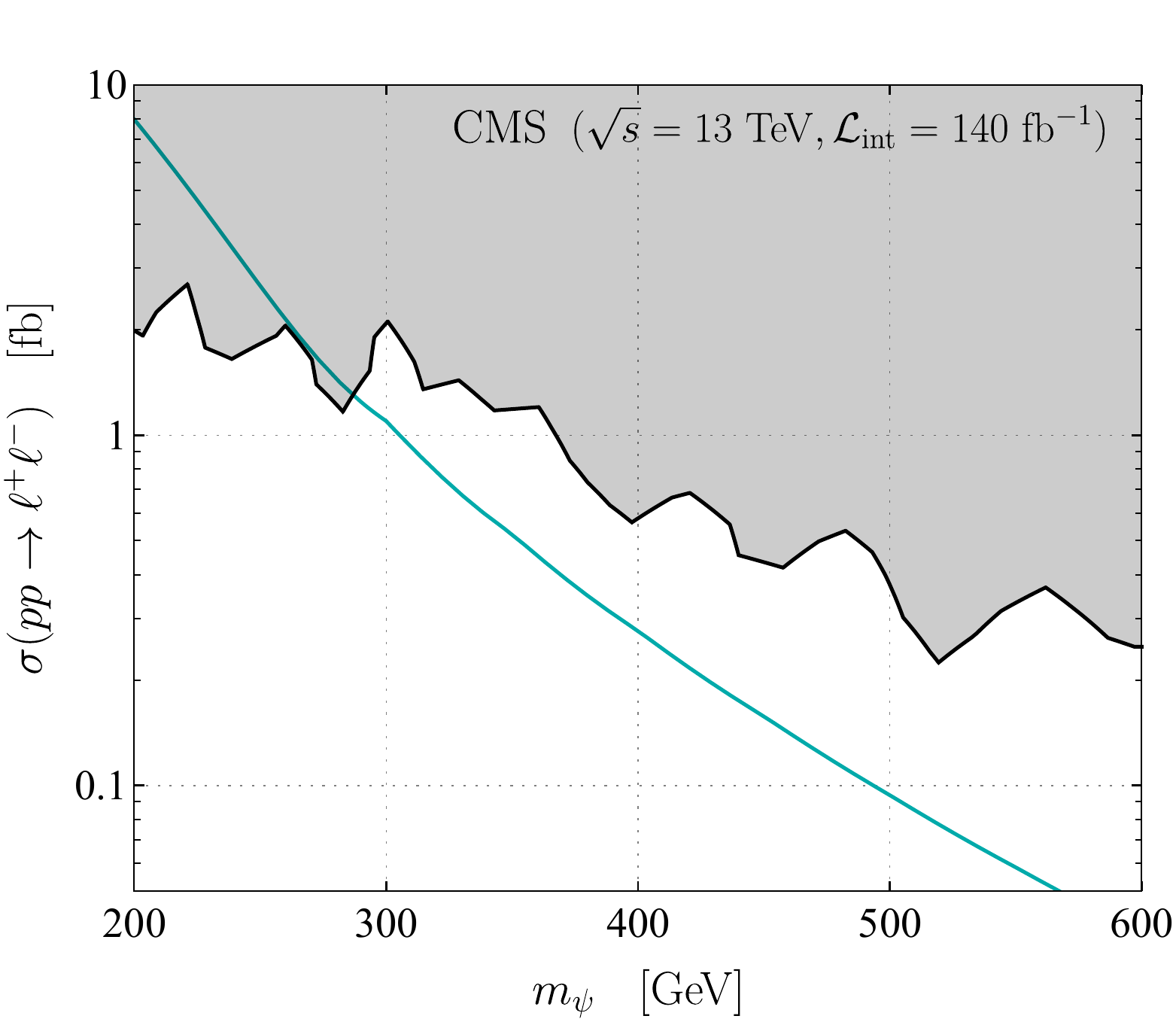}
  \end{tabular}
  \caption{The total cross-section (green curve) for the signal channel $pp \to \ell^+ \ell^-$, with $\ell=e,\mu$, as a function of $m_\psi$. The grey region is excluded by a $13\TeV$ CMS search with $140~\rm fb^{-1}$~\cite{CMS:2019tbu}.
  }
  \label{fig:psionium_cs}
\end{figure}

The cleanest channels for discovering the psionium states are the processes $pp\to\Psi^0\to\gamma\gamma$ and $pp\to\Psi^1\to\ell^+\ell^-$, which lead to a resonant peak in the diphoton and dilepton invariant mass spectra at the mass of the psionium, $m_{\Psi^{0,1}}\approx2m_\psi$. Since we found that the dilepton decay channel of the vector meson $pp\to\Psi^1\to e^+e^-,\mu^+\mu^-$ (depicted in left panel of \cref{fig:psionium_production_decay}) produces more stringent LHC limits, we focus on this process.\footnote{For the diphoton channel, we considered the limits imposed by the search~\cite{CMS:2018dqv}.} 
The partonic cross-section is given in the narrow-width approximation by
\begin{align} \label{eq:partonic_cs_psionium}
\hat \sigma (pp \to \ell^+ \ell^-) = \left[\hat \sigma (q\bar{q} \to \Psi^1) + \hat \sigma (gg \to \Psi^1 g)\right] \times \rm BR (\Psi^1 \to \ell^+ \ell^-)\,.
\end{align}
In \cref{app:formulae_psionium} we give the respective expressions for the partonic production cross-sections and the leading order decay rates. We used the package \texttt{RunDec}~\cite{Chetyrkin:2000yt} to take higher order loop-corrections in the running of the strong coupling into account (which is important since the process is highly sensitive to the value of the strong coupling). Finally, we convolve the partonic cross-sections with the parton distribution function (PDF) set~\texttt{PDF4LHC15\_nnlo\_mc}~\cite{Harland-Lang:2014zoa,Ball:2014uwa,Butterworth:2015oua}. The total cross-section is shown in \cref{fig:psionium_cs}, as function of $m_\psi$. Using the experimental upper bounds on the total cross-section for vector resonances given in ref.~\cite{CMS:2019tbu}, we find that this dilepton search excludes $m_{\psi} < 280\GeV$. Notice that this limit is independent of the exact values of the model parameters, as long as $\psi$ hadronizes.
 
\subsubsection{$\psi$ Pair Production}
\label{sec:openpsi}

We now turn to pair production of $\Psi_q$ bound states. These states predominantly decay through $\psi\to b\tau \chi_1$ or $\psi\to t\nu_\tau \chi_1$, since they are $\mathbb{Z}_2$--odd, leading to collider signatures $bb\tau\tau\chi_1\chi_1$, $bt\tau\nu_\tau\chi_1\chi_1$ or $tt\nu_\tau\nu_\tau\chi_1\chi_1$. Due to the large top mass, $\psi \to \chi_1 t \nu_\tau$ is typically kinematically forbidden for dark matter masses below 1 TeV, so we focus on $\psi \to \chi_1 b \tau$ decay channel (see the right panel of \cref{fig:psionium_production_decay}).  Since $m_\psi \gg m_q$, the light quark acts as a spectator and we can assume that $\Psi_q$ and $\psi$ share the same mass and decay width. Therefore, in a collider environment, we may neglect the presence of the light quark altogether.

ATLAS and CMS have performed searches for pair-produced scalar leptoquarks decaying into the $bb \tau \tau + E_T^{\rm miss}$ final state at $\sqrt{s} = 13\TeV$ with an integrated luminosity of $36.1\,{\rm fb}^{-1}$~\cite{CMS:2018txo,CMS:2018iye,ATLAS:2018uni,ATLAS:2019qpq}. Unfortunately, these searches do not set strong limits on the model since they rely on large $p_T$ selection cuts for the reconstructed $b$-jets and $\tau$-jets.  The $b$- and $\tau$-jet coming from $\psi$ are expected to be relatively soft, due to the compressed spectrum $m_\psi \gtrsim m_{\chi_1}$, and therefore fail to pass the basic selection criteria for these searches. Indeed, when recasting \cite{CMS:2018txo,CMS:2018iye,ATLAS:2018uni,ATLAS:2019qpq} we found very small selection efficiencies, and these searches provided no constraints on our parameter space. In the next section we propose a dedicated search strategy for this type of compressed spectrum signatures at the LHC.

However, the failure of the visible decay products of $\Psi_q$ to pass the typical LHC search cuts means that mono-$X$ searches may be sensitive. The picture is that the soft (undetected) $b\tau$ pairs and hard missing transverse energy recoil against hard electroweak or QCD initial-state radiation (ISR) that does pass the selection criteria. In ref.~\cite{Chakraborty:2018kqn}, the process $pp \to \mathcal{Q} \bar{\mathcal{Q}} + j$ is investigated, where $\mathcal{Q}$ is a heavy coloured particle that decays into a dark matter candidate (and SM particles) and $j$ is a hard (i.e., $p_T^j > 250\GeV$) ISR jet. The authors performed a recast of the ATLAS inclusive monojet search~\cite{ATLAS:2017bfj} and derive lower bounds on $m_{\mathcal{Q}}$ for benchmarks of different spin and colour representations. The results obtained at NLO QCD for a fermionic top-partner can be readily applied to our scenario. It is worth noticing here that the study fixes the mass gap $\Delta m = m_{\mathcal{Q}} - m_\text{DM} \lesssim 40\GeV$ in order to avoid transitions to multi-jet + MET signatures. The relevant part of the parameter space in our setup where this condition can be satisfied is $m_{\psi} < 400\GeV$, as can be seen in \cref{fig:relic-surface}. Coincidentally, the lower limit derived in \cite{Chakraborty:2018kqn} for the mass of the fermionic top-partner stands around the same value. Therefore, we conclude that inclusive monojet searches exclude $\psi$ masses below $400\GeV$ at $\sqrt{s}=13 \TeV$ with $\mathcal {L}=36.1 \,\rm fb^{-1}$ of data.

\subsubsection{Single $\psi$ Production}

It is possible to produce a single $\psi$ in association with the dark matter particle $\chi_1$ via $pp\to \chi_1\psi\tau$ (see \cref{tab:modes}). Such production processes would directly probe the leptoquark coupling, which must be large to accommodate the $B$-physics anomalies. Unfortunately, these are $2\to 3$ body transitions and are therefore heavily phase-space suppressed compared to the $2\to2$ processes. Moreover, the partonic initial state contains one $b$-quark, so this process is further suppressed by the small bottom PDF. While interesting, single $\psi$ production modes are not expected to set competitive limits on the parameter space. 

\section{Dedicated Search for $\psi$ Pairs}
\label{sec:new_search}

We propose a novel LHC search strategy for $\psi$ pair production, specifically $pp\to\psi\bar\psi\to b\bar b\tau^-\tau^+ \chi_1\chi_1$ assuming a compressed spectrum scenario, $m_\psi \gtrsim m_{\chi_1}$. We focus on the experimental signature $2j_b+\tau_h\tau_\ell+E^{\text{miss}}_T$, consisting of a pair of $b$-tagged jets ($j_b$), one $\tau$-tagged jet coming from a hadronically decaying $\tau$-lepton ($\tau_h$), one light lepton $e$ or $\mu$ coming from a leptonically decaying $\tau$-lepton ($\tau_\ell$), and a large amount of missing transverse energy ($E^{\text{miss}}_T$) mostly coming from the dark matter particles and $\tau$-neutrinos.\footnote{Signatures with two hadronic taus ($2j_b+2\tau_h+E^{\text{miss}}_T$) can also be considered. While this channel has a larger branching ratio, it suffers from important reducible backgrounds (e.g., fake taus) that are difficult to determine using MC simulations. Estimating the LHC sensitivity of a search strategy for this signal category is therefore beyond the scope of this paper.}
Notice that this signature is similar to the SLT (single-lepton trigger) category described in \cite{ATLAS:2018uni}. 
The irreducible backgrounds are dominated by $t\bar t$ pair production decaying into $t\to b\tau_h\nu$ and $t\to b\ell\nu$, and sub-leading contributions from Drell-Yan ($pp\to\tau^+_h\tau^-_\ell\ +\ $jets) and diboson production ($pp\to ZZ\to\tau^+_h\tau^-_\ell j_bj_b$). The main reducible backgrounds consist of processes leading to objects misidentified as $\tau$-jets. These fake taus arise from any process generating a prompt light lepton in association with jets and $b$-jets, with one of the QCD jets mistagged as a hadronic $\tau$-lepton, $j\to\tau_h$. In order to simulate the signal and background samples, we first implemented in {\tt FeynRules} \cite{Alloul:2013bka} the interaction Lagrangian \cref{eq:LU} and generated the UFO model \cite{Degrande:2011ua} for the Monte Carlo event generator. The parton level event samples were simulated using {\tt MadGraph5} \cite{Alwall:2014hca} and both showering and hadronization were performed in {\tt Pythia8} \cite{Sjostrand:2014zea}. Jets were clustered using {\tt FastJet3} \cite{Cacciari:2011ma}, while object reconstruction and detector effects were simulated with {\tt Delphes3} \cite{deFavereau:2013fsa}. 

For the signal process, we simulated event samples for different points in the ($m_{\chi_1},m_U$) mass plane with the mass splitting $\Delta_\psi$ fixed by the observed relic abundance constraint, as shown in \cref{fig:relic-surface}. Moreover, the leptoquark couplings were fixed by \cref{eq:gUmU,eq:minCoup} in order to satisfy the $B$--anomalies. For the background processes, in our analysis we only included the leading $t\bar t$ process given that the other irreducible backgrounds are sub-leading. Since the $j\to\tau_h$ mistagging efficiency are not easy to estimate without using data-driven techniques, we did not include these backgrounds in our analysis. Note that such backgrounds, while typically smaller than the leading $t\bar t$ process, are a non-negligible background that must be included in a more thorough analysis.

In the following we describe two analyses.  One is a cut-based search using a single high-level observable $R_T$ (defined below) as a signal/background discriminator, while the other is a multi-variate analysis using a BDT classifier based on multiple low-level and high-level observables.

\subsection{Event Selection and Key Observables}
\label{sec:event-selection}

We now describe the basic event selections used in our analyses. Jets were clustered using the anti-$k_T$ algorithm with the standard narrow cone radius of $R=0.4$ and candidate leptons $\ell=e,\mu$ were selected if they passed the relative isolation requirements $I_{\text{rel}}=0.05$ within an isolation cone of radius $R_{\text{iso}}=0.2$. In order to improve the signal acceptance, it is important to relax the transverse momentum cuts as much as possible for any of the visible reconstructed objects since these are expected to be soft, due to the compressed spectrum. In our analysis we required $p_T(j)>20\GeV$, $p_T(\tau_h)>20\GeV$, and $p_T(\ell)>5\GeV$ for jets, $\tau$-jets and isolated leptons, respectively. The rapidity selections were taken to be the standard ones: $|\eta(j)|<2.5$, $|\eta(\tau_h)|<2.3$, and $|\eta(\ell)|<2.5$ with isolated electrons removed from the end-cap region ($1.37 < |\eta(e)|< 1.52$). For the $b$-tagging and $\tau$-tagging efficiencies we used the default values of the ATLAS card in {\tt Delphes3}. Finally, events are selected if they contain two or more jets, of which at least one must be $b$-tagged, exactly one $\tau$-jet and exactly one light lepton $\ell$, such that the $\tau_h \ell$ pair has opposite electric charge.

 \begin{figure}[t]
  \centering
  \begin{tabular}{cc}
  \includegraphics[width=1\textwidth]{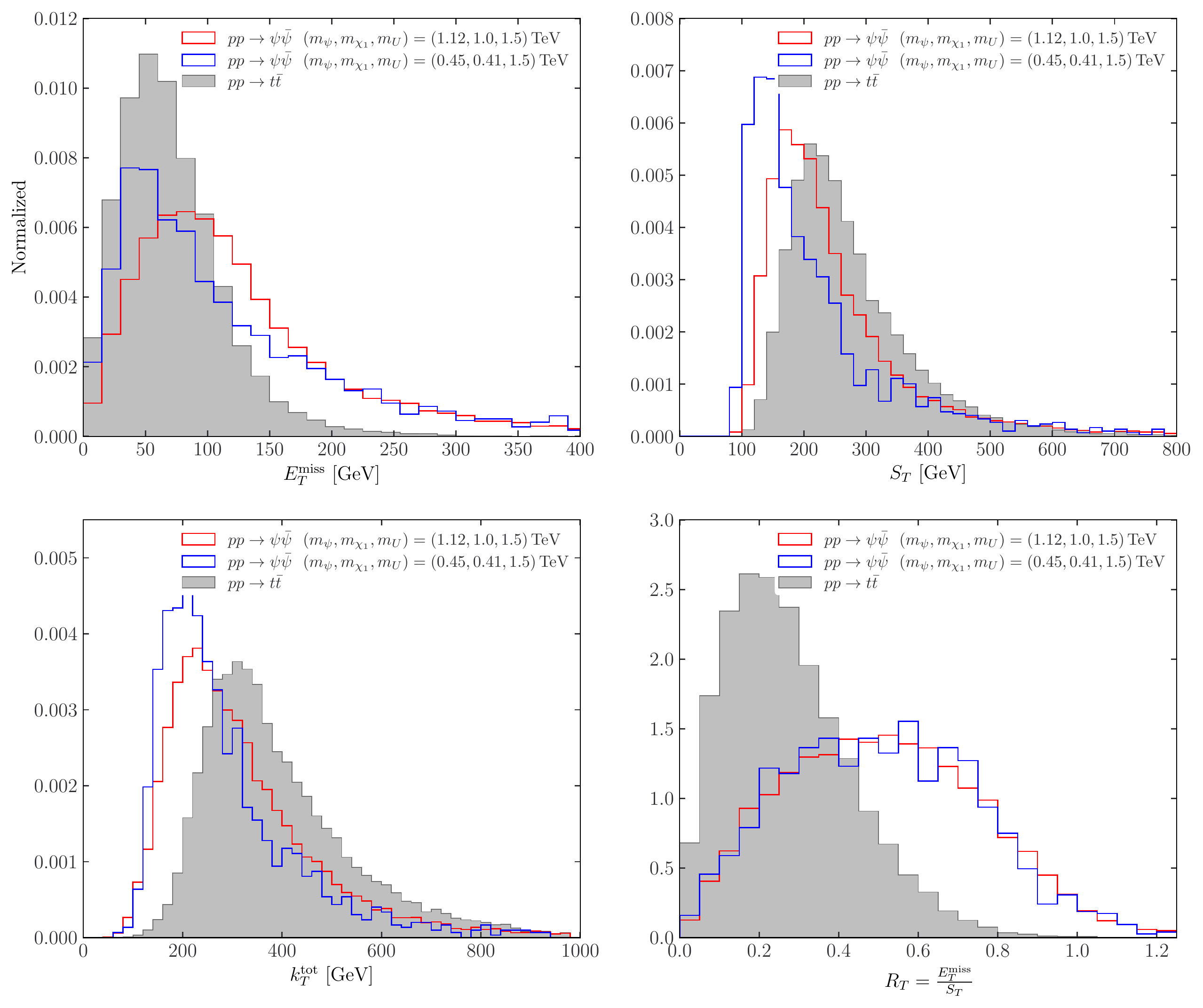}
  \end{tabular}
  \caption{Normalised $E^{\text{miss}}_T$, $S_T$, $k_T^{\text{tot}}$ and $R_T$ event distributions for two benchmark signal processes (red, blue) and for the $t\bar t$ background (grey). See main text for details. } 
  \label{fig:discriminators}
\end{figure}

After these basic selection cuts, we looked into different possible observables capable of distinguishing between the signal and the $t\bar t$ background. The two simplest discriminating quantities for this task are the missing transverse energy, $E^{\text{miss}}_T$, and the total amount of visible transverse momentum, defined as 
\begin{equation}
S_T\equiv p_T(j_{1})+p_T(j_{2})+p_T(\tau_h)+p_T(\ell) \,,
\end{equation}
i.e., the scalar sum of the transverse momenta of the leading and sub-leading jets (with at least one being $b$-tagged), the $\tau$-jet and the lepton. On one hand, because of the presence of the dark matter particles and neutrinos, signal events will tend to have more missing energy than $t\bar t$ events. On the other hand, because of the compressed spectrum, the visible energy in signal events will tend to be softer than in $t\bar t$ events. In the first two upper panels in \cref{fig:discriminators} we show these two normalised distribution for two signal benchmarks values $(m_\psi,m_{\chi_1},m_U)=(1.12,1.0,1.5)\TeV$ (red) and $(m_\psi,m_{\chi_1},m_U)=(0.45,0.41,1.5)\TeV$ (blue) that satisfy the dark matter relic abundance requirement, and the $t\bar t$ background (shaded grey). 

In order to examine the relevance of angular information between final state objects, we also looked into the 2-point energy correlation functions. These observables are defined as
\begin{equation}\label{eq:summed_obs}
\mathcal{C}(x_1,\cdots,x_n)=\sum_{i<j}^n p_{T_i}p_{T_j} d^2_{ij}\,,
\end{equation}
for a set $x_1,\cdots,x_n$ of high-level reconstructed objects in the event.\footnote{ $n$-point energy correlation functions~\cite{Larkoski:2013eya} are more commonly measured between the individual constituents of a single jet.} 
Here $d_{ij}$ is the angular distance between two objects $x_i$ and $x_j$, with transverse momenta $p_{T_i}$ and $p_{T_j}$, respectively. We considered two possible distance functions: the azimuthal distance $\Phi_{ij} \equiv\sqrt{2(1-\cos\Delta\phi_{ij})}$ and the plane distance $R_{ij} \equiv\sqrt{(\Delta y_{ij})^2+(\Delta\phi_{ij})^2}$, where $\Delta y_{ij} =y_i-y_j$, $\Delta \phi_{ij}=\phi_i-\phi_j$ are determined by the rapidity $y_i$ and azimuth angle $\phi_i$ of the object $x_i$. Notice that for $d_{ij}=\Phi_{ij}$ in \cref{eq:summed_obs}, each term reduces to the (squared) transverse mass $M^2_T$ of a pair of objects, and the observable corresponds to the square of the total transverse mass $M_T^{\text{tot}}$ of the group of objects $x_1,\cdots,x_n$. If instead we fix $d_{ij}=R_{ij}$, then each term in \cref{eq:summed_obs} corresponds to the $k_T$-distance between the pairs of objects. We denote this observable $k_T^{\text{tot}}$ and refer to it as  the total $k_T$-distance between $x_1,\cdots,x_n$. In \cref{fig:discriminators} (left lower panel) we show as an example the $k_T^{\text{tot}}$ observable computed for the set of visible final states $\tau_h$, $\ell^\pm$, $j_1$ and $j_2$, for signal and background. Notice that these observables containing pairwise angular correlations do not substantially improve the signal separation compared to observables such as $S_T$ without angular information (right upper panel).\footnote{An improvement is more apparent for the $(m_\psi,m_{\chi_1},m_U)=(1.12,1.0,1.5)$ benchmark (red) for which the visible final states are expected to be slightly harder and therefore more challenging to distinguish from the $t\bar t$ background.}   
$M_T^{\text{tot}}$, based on $d_{ij}=\Phi_{ij}$, gives very similar results.

Individually, none of the observables mentioned so far lead to a good separation between signal and background. Interestingly, much better discriminators are obtained by combining pairs of these observables. We found that the ratio between the invisible and visible total transverse momentum of the event, defined by
\begin{equation}\label{eq:RT}
R_{T}\equiv \frac{E^{\text{miss}}_T}{S_T}\,,
\end{equation}
considerably improves the signal separation compared to the individual $E_T^\text{miss}$ and $S_T$ observables. This can be seen in the lower right panel of \cref{fig:discriminators}, where we show the normalised $R_T$ distributions. Moreover, the shapes of the distribution for the signal process are independent of the particle masses $m_\psi,\, m_{\chi_1}$, which is not the case for $E_T^\text{miss}$, $S_T$, $M_T^\text{tot}$ or $k_T^\text{tot}$. Finally, we also checked other ratios similar to \cref{eq:RT} by replacing the denominator with $M_T^\text{tot}$ and $k_T^\text{tot}$. These ratios were found to have comparable performance to, and the same properties as, $R_T$.  We chose to cut on $R_T$ in our search, and we present the results in \cref{sec:results}.

\subsection{Multivariate Analysis}

 \begin{figure}[t]
  \centering
  \includegraphics[width=1\textwidth]{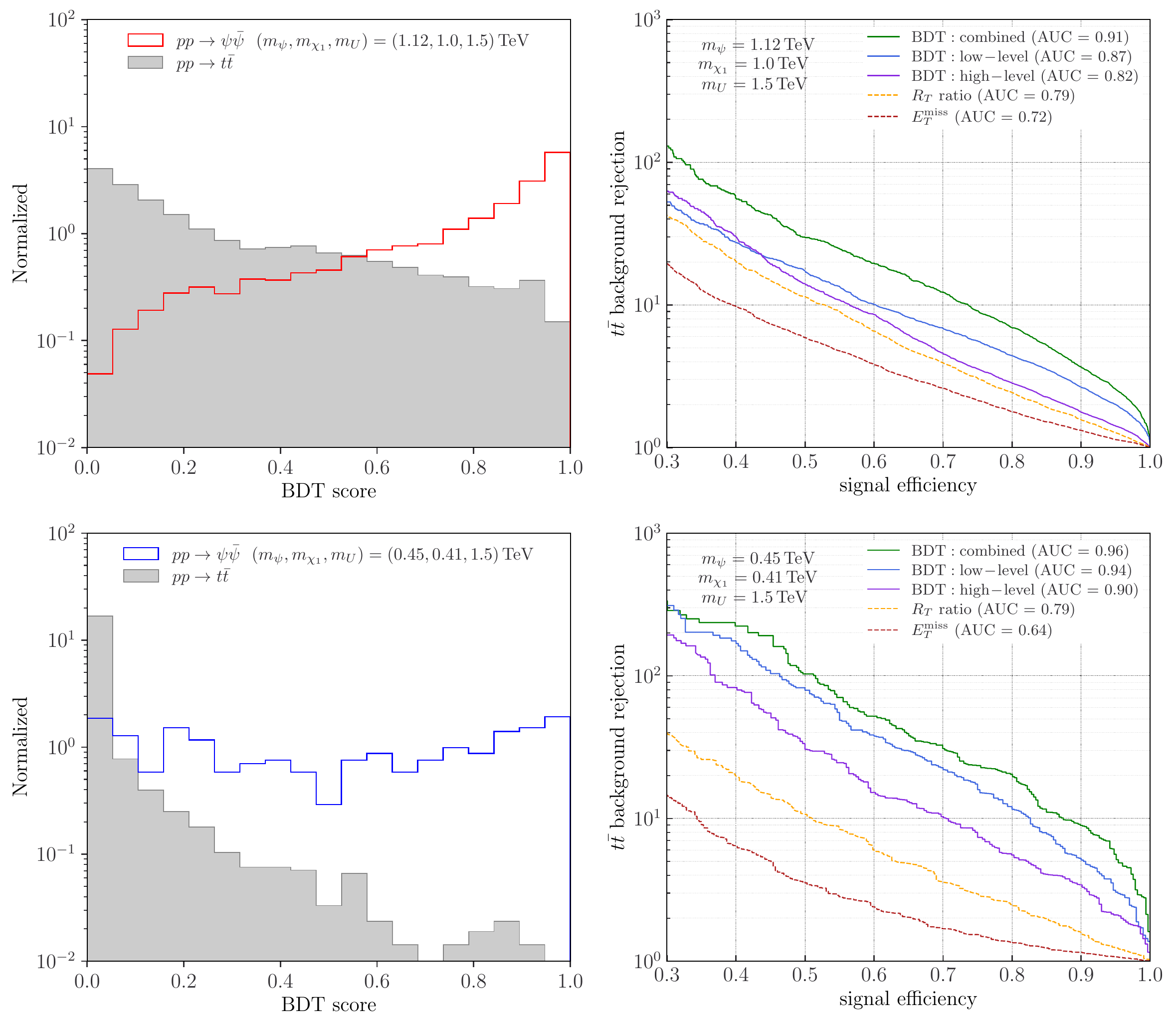}
  \caption{Left panels: BDT scores for the two different benchmark points showing signal vs background discrimination. Right panels: ROC curves for the same benchmarks showing the $t\bar t$ background rejection rate vs signal efficiency for three different BDT classifiers based on only high-level observables (purple curve), low-level observables (blue curve) and both high-level and low-level observables (green curve). For comparison, we also include the cut-based ROC curves for the $R_T$ ratio (dashed yellow) and $E_T^\text{miss}$ (dashed red). Better performance is achieved by ROC curves stretching towards the upper-right corner and AUC values closer to $1$.
  }
  \label{fig:Rocs}
\end{figure}

Given that a combination of observables ($R_T$) performs better than the individual ones ($E^{\text{miss}}_T$ and $S_T$), we also implemented a multivariate search using a BDT classifier. After imposing the same event selections as in \cref{sec:event-selection}, a BDT classification score was extracted using low-level and high-level observables as input variables. The low-level observables consist of the $p_T$ and $\eta$ distributions of the relevant reconstructed objects ($j_{1,2}$, $\tau_h$ and $\ell$) while the high-level observables consist of $E^{\text{miss}}_T$, $S_T$, $M_T^\text{tot}$, $k_T^\text{tot}$ and the $R_T$ ratio defined in \cref{eq:RT}. 

We tested three different BDT setups: (i) the {\it high-level} BDT trained with only high-level observables, (ii) the {\it low-level} BDT trained with only low-level observables, and (iii) the {\it combined} BDT trained with both high-level and low-level observables. For each both signal benchmarks, we prepared an event sample with $\sim\!45$\,K labelled events and a signal to background ratio of $s/b\approx1.2$. The samples were then split into $80\%$ training and $20\%$ testing sub-samples. The training was performed with {\tt XGBoost} \cite{Chen2016} using $100$ trees with a maximum depth of $6$, learning rate of $\eta=0.1$ and a binary logistic for the objective function. 

The resulting scores extracted from the combined BDT classifier, trained with the hyperparameters described above, is shown in the first column in \cref{fig:discriminators}, for two signal benchmark mass points. A very clear separation between each benchmark signal and the $t\bar t$ background is obtained for this classifier. Furthermore, we measured the performance of the BDT classifiers using the receiving operating curves (ROC) displayed in the second column in \cref{fig:Rocs}. There, one can observe that the three BDTs perform well (solid curves), achieving $t\bar t$ rejection rates  above $10$ for a signal efficiency of $50\%$ and area-under-the-curve (AUC) values above $\sim0.8$. Overall, the low-level BDT (blue curve) performs better than the high-level BDT (purple curve), but the best classifier is the combined BDT (green curve) with AUCs above $0.9$. The three BDTs considerably outperform the cut-based single discriminators (dashed curves). In this case the best cut-based observable was the $R_T$ ratio which produces moderate background rejection with an AUC close to $0.8$.

These results demonstrate that a BDT classifier trained on both low-level and high-level observables can improve the $t \bar t$ background rejection rate by at least a factor of two compared to the cut-based strategy based on the best high-level single observable, the $R_T$ ratio.

\subsection{Results}
\label{sec:results}

A set of signal event samples for different values of $(m_{\chi_1},m_U)$ were generated covering the parameter space region $200\GeV<m_{\chi_1}<1000\GeV$ and $1500\GeV<m_U<5000\GeV$, with a rectangular grid with spacing $(\delta m_{\chi_1},\delta m_U)=(50,200)\GeV$. In order to get statistically significant MC samples (especially in the tails of the $R_T$ distribution) we simulated around $1$\,M signal events for each mass point of the grid. For the background, we generated $500$\,K $t\bar t$ events. 

 \begin{figure}[t!]
  \centering
  \includegraphics[width=0.8\textwidth]{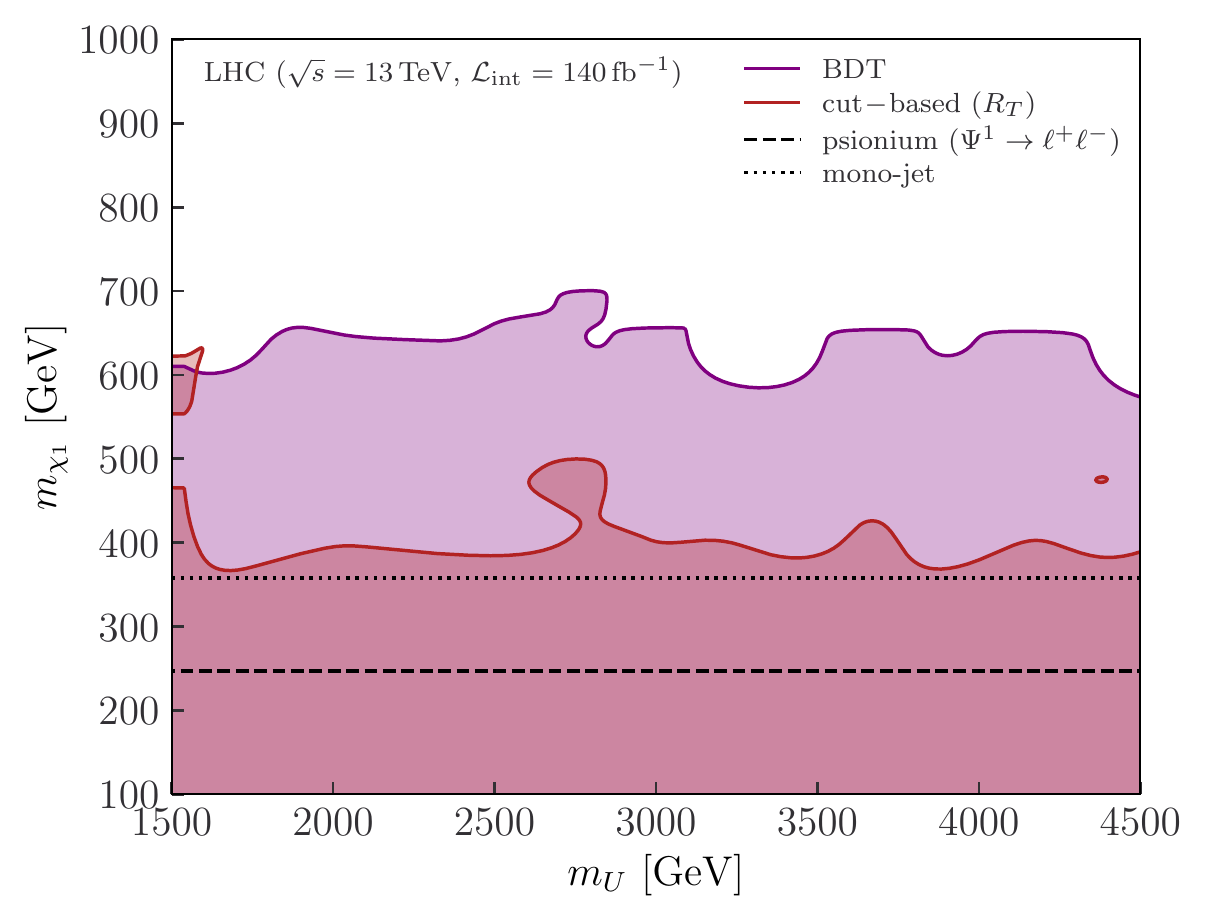}
  \caption{Exclusion limits at $95\%$ CL for the cut-based analysis based on the $R_T$ observable (red region) and the multivariate analysis based on a combination of low-level and high-level observables (purple region).  For comparison, the psionium and monojet limits are also shown.
  }
  \label{fig:limits}
\end{figure}

For the cut-based strategy, after performing the event selections for each sample, we performed a statistical analysis using the binned $R_T$ distribution in the interval $[0.5,1.0]$ with a bin step of $0.1$ including overflow in the last bin. Events were estimated with run-II integrated luminosity of $\mathcal{L}_{\text{int}}=140\,\text{fb}^{-1}$ and a centre-of-mass energy of $\sqrt{s}=13\TeV$. In each $R_T$ bin, we fixed a 10\% uncertainty for the $t\bar t$ background, which is approximately the same uncertainty that was estimated for the background $S_T$ distributions in ref.~\cite{ATLAS:2019qpq}. For the multivariate analysis, we trained the combined BDT on each point of the rectangular grid using the hyperparameters described above but with 50\% training and 50\% testing sub-samples (in order to retain enough MC events in the testing sample).  We then used the binned score of the BDT classifier with $20$ equidistant bins over the unit interval. Following ref.~\cite{ATLAS:2019qpq}, we assumed a $10\%$ background uncertainty for the BDT score. For each of the two search strategies, the expected exclusion limits were extracted at each point in the $m_{\chi_1}$--$m_U$  plane using the asymptotic approximation of the CL$_s$ criteria \cite{Read:2002hq,Cowan:2010js} with the profile likelihood ratio as test statistic. This statistical model was implemented using the {\tt pyhf} python package \cite{pyhf,pyhfjoss}. A mass point in the $m_{\chi_1}$--$m_U$ plane is excluded at $95\%$ confidence level (CL) if $CL_s<0.05$ is satisfied. In \cref{fig:limits} we present the final upper exclusion regions in the $m_{\chi_1}$--$m_U$ plane given by the red area for the cut-based analysis and the purple area for the multivariate analysis. For comparison, we have also included the model-independent limits from the psionium production $pp\to\Psi^1\to\ell^+\ell^-$ (black dashed), as well as the expected reach of a monojet search (black dotted). We can see that both the cut-based and multivariate dedicated searches will give stronger limits that these recasted limits. Moreover, the multivariate analysis significantly outperforms the cut-based limits.

\section{Conclusions}
\label{sec:conclusions}

In this work we have investigated coannihilating dark matter, motivated by the vector leptoquark explanation of the $B$-physics anomalies. Assuming that the leptoquark is a gauge boson of a spontaneously broken gauge symmetry, and that dark matter is a fermion contained in a multiplet of that symmetry, the dark sector will also contain a coloured coannihilation partner.  Furthermore, introducing UV-motivated dimension-5 operators leads to a Majorana dark matter candidate which is similar in mass to the coannihilation partner. 

We determined the mass splitting between the Majorana dark matter particle $\chi_1$ and its coannihilation partner $\psi$ required to reproduce the observed dark matter relic abundance via thermal freeze-out (see \cref{fig:relic-surface}). Interestingly, processes with a leptoquark mediator are significantly more efficient than those with a $Z'$ mediator.  We found that the relic abundance can be satisfied for dark matter masses up to $\sim10\TeV$ and for leptoquark masses in the motivated range from 1.7--$10\TeV$.  We then analysed the phenomenology of the model for the slice of parameter space which reproduces the observed relic abundance.  Direct and indirect constraints are negligible as the dark matter candidate is a Majorana fermion, so collider experiments are the best probe of the parameter space.

Since leptoquark searches have been well studied and searches for the dark matter candidate in our setup are challenging, we focused on the coloured coannihilating partner, which can be pair produced with a relatively large cross-section at the LHC. The coannihilating partner lifetime is long enough for it to hadronize, forming both `psionium' and `open-psi' bound states.  Psionium can decay via electroweak interactions into dileptons, leading to a lower bound of $280\GeV$ on the mass of the coannihilating partner. Open-psi predominantly decays to dark matter along with a $b$ quark and a $\tau$ lepton via an off-shell $U_1$ leptoquark.  Due to the compressed spectrum, the $b$ quarks and $\tau$ leptons will be soft and may not pass selection cuts.  We found that existing searches for $bb \tau \tau + E_T^{\rm miss}$ do not place constraints on the relevant parameter space.  However, monojet searches are sensitive and currently exclude coloured coannihilation partner masses below $400\GeV$.  

We proposed a new search strategy centred around a new observable: the ratio between visible and missing transverse energy of the process. We performed both a cut-based and a multi-variate analysis that directly probe the mass of the coloured partner, assuming LHC run-II luminosities. Using the relic surface, we determined expected limits on the dark matter mass: $m_{\chi_1} \lesssim 400\GeV$ and $m_{\chi_1} \lesssim 600\GeV$ for the two analyses, respectively.  We therefore conclude that a multi-variate analysis would significantly improve on current limits and could probe a significant portion of viable parameter space.  

Our final results are summarised in \cref{fig:limits}. Although this analysis was motivated by dark matter and the $B$-physics anomalies, we emphasise that (i) bound state formation and decay could be relevant for other coannihilating scenarios with coloured partners, and (ii) the ratio of visible to missing transverse energy could be a powerful discriminator in scenarios with compressed spectra and long-lived invisible particles. 

Finally, while our results are encouraging, a more detailed analysis performed by the LHC experimental collaborations would provide a more robust estimation of the limits.  For instance, including the $2j_b + \tau_h\tau_h + E^{\text{miss}}_T$ signal category would strengthen the limits, and better accounting for mistagging rates and reducible and irreducible backgrounds would improve the accuracy of the limits.  Furthermore, it would be necessary to carefully consider the trigger requirements for processes with soft final state objects, e.g., the possibility of triggering on our new observable $R_T$ could be investigated.

\section{Acknowledgements}
\label{sec:acknowledgements}

The authors would like to thank Ben Kilminster, for asking us questions that we hope this work goes some way to answering, Gino Isidori, for collaboration in the early stages of the project and Ben Stefanek, Julie Pages, Vinicius Mikuni and Arne Reimers for useful conversations. M.J.B.\ was supported by the Australian Research Council and by the Swiss National Science Foundation (SNF) under contract 200021-159720. S.T. acknowledges support by MIUR grant PRIN 2017L5W2PT. D.A.F has received funding from the European Research Council (ERC) under the European Union’s Horizon 2020 research and innovation programme under grant agreement 833280 (FLAY), and by the Swiss National Science Foundation (SNF) under contract 200021-175940.

\crefalias{section}{appendix}
\begin{appendices}
\section{4321 models}
\label{app:4321}
\setcounter{equation}{0}
\renewcommand\theequation{A.\arabic{equation}}

We now briefly describe a SM extension that gives rise to a TeV~scale $U_1\sim({\bf3,1},2/3)$ vector leptoquark as a gauge boson with non-universal couplings to quarks and leptons. These so-called 4321 models~\cite{DiLuzio:2017vat,Bordone:2017bld,Greljo:2018tuh,Bordone:2018nbg,DiLuzio:2018zxy,Cornella:2019hct,Fuentes-Martin:2020bnh,King:2021jeo} are defined by the gauge group 
\begin{equation}
    G_{4321}=SU(4)\times SU(3)_{c^\prime}\times SU(2)_L\times U(1)_{Y^\prime}\,,
\end{equation}
where $SU(2)_L$ is identified with SM weak isospin and both colour and hypercharge are diagonally embedded as $SU(3)_{c}\times U(1)_{Y}\subset SU(4)\times SU(3)_{c^\prime}\times U(1)_{Y^\prime}$. The breaking $G_{4321}\to G_{\rm SM}$ is typically induced by at least two scalar fields, $\Omega_1\sim({\bf\bar4,1,1},-1/2)$ and $\Omega_3\sim({\bf\bar 4,3,1},1/6)$, each developing a vev around the TeV scale.  A Higgs doublet $H\sim({\bf 1,1,2},-1/2)$ is necessary for electroweak symmetry breaking. The resulting gauge sector (in addition to the SM gauge bosons) consists of a colour octet $G^\prime\sim({\bf8,1},0)$, a colour singlet $Z^\prime\sim({\bf1,1},0)$ and the leptoquark $U_1\sim({\bf3,1},2/3)$, all with masses above $1\TeV$.
 
There are two common 4321 implementations distinguished by the charge assignments of the SM matter fields:
\begin{itemize}
    \item {\em (i) Standard 4321}: All three would-be SM fermion families $q^{i}_L$, $\ell^{i}_L$, $u^{i}_R$, $d^{i}_R$, $e^{i}_R$, are singlets under $SU(4)$ and have the usual SM charges under $SU(3)_{c^\prime}\times SU(2)_L\times U(1)_{Y^\prime}$. These fields therefore do not couple directly to any of the massive $SU(4)$ gauge bosons. Three vector-like fermions $\Psi^{i}_{L,R}\sim({\bf 4,1,2},0)$ are introduced in order to generate non-universal effective couplings between the SM left-handed fermions and the $U_1$ gauge leptoquark via fermion mixing, which arises from the Yukawa interactions $\bar \ell^{i}_L\Omega_{1}\Psi^{j}_R$ and $\bar q^{i}_L\Omega_{3}\Psi^{j}_R$ after spontaneous symmetry breaking. The matter content is shown in the first block for $i=1,2,3$ in \cref{tab:4321}.
    \item {\em (ii)  Flavoured 4321}: In this case the would-be third family quarks and leptons are unified into $SU(4)$ mulitplets $\Psi^3_L=(q^3_L,\ell^3_L)^T$, $\Psi^{3+}_R=(t_R,\nu_R)^T$ and $\Psi^{3-}_R=(b_R,\tau_R)^T$ and couple directly to the $U_1$ leptoquark. The first two would-be SM generations are $SU(4)$ singlets and couplings to $U_1$ are induced via fermion mixing with two vector-like fermions $\Psi^{1,2}_{L,R}$, as in the standard 4321. The matter content is shown in the first and second blocks for $i=1,2$ in \cref{tab:4321}.
\end{itemize}

\begin{table}[t]
  \centering
    \begin{tabular}{@{\hspace{1em}}c@{\hspace{2em}}c@{\hspace{2em}}c@{\hspace{2em}}c@{\hspace{2em}}c@{\hspace{1em}}}
    \hline\hline
     Fields & $SU(4)$ & $SU(3)_{c^\prime}$  & $SU(2)_{L}$  &$U(1)_{Y^\prime}$   \\
    \hline
     {$q^i_L$}       & ${\bf 1}$   & ${\bf 3}$   & ${\bf 2}$   & $1/6$    \\ 
     {$\ell^i_L$}    & ${\bf 1}$   & ${\bf 1}$   & ${\bf 2}$   & $-1/2$  \\
     {$u^i_R$}       & ${\bf 1}$   & ${\bf 3}$   & ${\bf 1}$   & $2/3$    \\ 
     {$d^i_R$}       & ${\bf 1}$   & ${\bf 3}$   & ${\bf 1}$   & $-1/3$  \\ 
     {$e^i_R$}       & ${\bf 1}$   & ${\bf 1}$   & ${\bf 1}$   & $-1$    \\ 
     {$\Psi^i$}      & ${\bf 4}$   & ${\bf 1}$   & ${\bf 2}$   & $0$     \\ 
    \hline
     {$\Psi^{3+}_R$}  & ${\bf 4}$   & ${\bf 1}$   & ${\bf 1}$   & $1/2$   \\ 
     {$\Psi^3_L$}    & ${\bf 4}$   & ${\bf 1}$   & ${\bf 2}$   & $0$      \\ 
     {$\Psi^{3-}_R$}  & ${\bf 4}$   & ${\bf 1}$   & ${\bf 1}$   & $-1/2$  \\
     \hline\hline
    \end{tabular}
\caption{ Matter sector of the standard 4321 model (first block with $i=1,2,3$) or the flavoured 4321 model (first and second blocks with $i=1,2$).}
\label{tab:4321}
\end{table}

Besides the minimal field content described above, more scalar or fermion fields are sometimes necessary to satisfy additional phenomenological requirements. For instance, symmetry breaking scalars $\Omega_{15}$ transforming in the adjoint representation $\bf 15$ of $SU(4)$ can be included in order to induce mass splittings between gauge bosons and fermion components \cite{DiLuzio:2018zxy,Cornella:2019hct}. These fields will couple to $SU(4)$ fundamentals as $\bar\Psi\Omega_{15}\Psi$ and a vev along the $T^{15}=\rm{diag}(1,1,1,-3)/\sqrt{6}$ generator will lead to a mass splitting of order $\langle\Omega_{15}\rangle$ between the coloured and colourless components of $\Psi$. Other 4321 matter extensions require the presence of fermion singlets $({\bf 1,1,1},0)$ that give rise to light Majorana neutrinos through the inverse seesaw mechanism \cite{Greljo:2018tuh, Fuentes-Martin:2020pww}. Finally, the 4321 models can be viewed as the low-energy limit of a more fundamental theory, such as the Pati-Salam cube model (PS$^3$) \cite{Bordone:2018nbg} (which can be embedded into a warped $5$D construction \cite{Allwicher:2020esa,Fuentes-Martin:2020pww}), the twin Pati-Salam model \cite{King:2021jeo} or strongly coupled models with extended hypercolour \cite{Fuentes-Martin:2020bnh}.

\section{Partial Width Formulae}
\label{app:lq_widths}
\setcounter{equation}{0}
\renewcommand\theequation{B.\arabic{equation}}

The tree-level partial decay width of the $U_1$ vector leptoquark into massive SM quarks and leptons is
\begin{align}
\Gamma_{U_1\to q_i\ell_j}&=g_U^2\left(|\beta_L^{ij}|^2+|\beta_R^{ij}|^2\right)\frac{ \lambda(m_U^2,m_q^2,m_\ell^2)}{48\pi\,m_U}\,\left[1-\frac{m_q^2+m_\ell^2}{2m_U^2}-\frac{(m_q^2-m_\ell^2)^2}{2m_U^4} \right]\,,
\end{align}
while the partial width into final state dark vector-like fermions is
\begin{align}
\Gamma_{U\to \psi\chi}&=g_U^2|\beta_D|^2\frac{ \lambda(m_U^2,m_\psi^2,m_{\chi_1}^2)}{24\pi\, m_U}\,\left[1-\frac{m_\psi^2+m_{\chi_1}^2-6m_\psi m_{\chi_1}}{2m_U^2}-\frac{(m_\psi^2-m_{\chi_1}^2)^2}{2m_U^4} \right]\,,
\end{align}
where we have used K\"allen's function $\lambda(x,y,z)\equiv\sqrt{x^2+y^2+z^2-2xy-2yz-2zx}$.

In the limit where the $\mathbb{Z}_2$--odd particles are significantly heavier than the SM particles and lighter than the leptoquark, the $\psi$ width into $\chi_1 b \tau$ is
\begin{align}\label{eq:decay_width}
\Gamma_{\psi \to \chi_1 b \tau} = \frac{g_U^4 |\beta_{d}|^2(|\beta_L|^2+|\beta_R|^2)m_{\psi}^5}{3072 \pi^3 M_U^4}\ f(m_{\chi_1} / m_{\psi})~ + \mathcal{O}\left(M_U^{-6} \right) \,,
\end{align} 
where $f(x) \equiv 1  - 2 x - 8 x^2 - 18 x^3 + 18 x^5+ 8 x^6 + 2 x^7 - x ^8  - 24 x^3  (x^2 + x + 1) \log x \to 0$ as $x\to 1$. In case of a compressed spectrum $m_{\chi_1}\sim m_\psi$, $f(m_{\chi_1}/m_\psi)\ll1$ and the width is significantly suppressed.

\section{Bound State Formation}
\label{app:hadronization}
\setcounter{equation}{0}
\renewcommand\theequation{C.\arabic{equation}}

To describe the bound states we use non-relativistic QCD with a modified-hydrogenic model of a single-gluon exchange potential~\cite{Fabiano:1997xh}. For coloured particles of mass $m \gg \Lambda_{\rm QCD}$, the potential takes the Coulombic form,
\begin{equation}
V(r) = - C \frac{\bar\alpha_s}{r}\,,
\end{equation}
where $C$ is a colour factor.  For a colourless bound state $C$ is simply the quadratic Casimir of the constituent particles (i.e., $C=4/3$ for the psionium). We define 
\begin{equation}
\bar \alpha_s = \alpha_s(r_B^{-1})\,,
\end{equation}
as the running strong coupling evaluated at the scale of the average distance $r_B$ between the two constituents. This distance is of the order of the Bohr radius $a_0 = 1/(C \bar\alpha_s \mu)$, where $\mu$ is the reduced mass of the system.  More precisely, for an $S$-wave ground state $r_B=\sqrt{3} a_0$. 

Bound states form if the revolution time, $t_R$, is larger than the lifetime of the coloured particles.  Combining the ground state energy
\begin{equation}\label{eq:energy1}
E_1 = - \frac{C^2 \bar \alpha_s^2 \mu }{2}\,.
\end{equation}
with the virial theorem yields
\begin{equation}\label{eq:tR}
    t_R= \frac{2 \pi} { C^2 \bar \alpha_s^2 \mu}\,.
\end{equation} 

\section{Production and Decay Rate of Psionium $\Psi^1$}
\label{app:formulae_psionium}
\setcounter{equation}{0}
\renewcommand\theequation{D.\arabic{equation}}

To determine the production cross-sections and decay widths of $\Psi^1$ we follow a generalisation of the results that apply to quarks and quarkonia~\cite{Kats:2012ym}. The partial width for $\Psi^1$ decaying into a final state $X$ can be written as
\begin{equation} 
\Gamma (\Psi^1 \to X) \to \Gamma (\psi \bar\psi \to X) |\Psi(0)|^2\,,
\end{equation}
where $\Psi(0)$ is the wavefunction at origin for the ground state of the bound system and $\Gamma (\psi \bar\psi \to X)$ is the corresponding process with a free pair of $\psi$ and $\bar\psi$ in the initial state. In the  modified-hydrogenic model, we obtain the following wavefunction 
\begin{equation} 
|\Psi(0)|^2 = \frac{1}{\pi} \left( \frac{C M \bar \alpha_s}{4} \right)^3\,,
\end{equation}
where $M$ is the psionium mass.

The leading production mechanisms of $\Psi^1$ are:
\begin{enumerate}[i)]
	\item \textbf{Electroweak production from $q\bar{q}$.} In the approximation that the psionium mass $M$ is much larger than the $Z$ boson mass, we may write the electroweak cross-section via an $s$-channel photon or $Z$ boson as
\begin{equation}
		\sigma(q\bar{q} \to \Psi) = \frac{\pi^2}{108}\, D_R C_R^3 Q^2\,
\frac{\alpha^2 \bar\alpha_s^3}{\cos^4\theta_W} \left(17\sum_{q=u,c} + 5\sum_{q=d,s,b}\right)\frac{\mathcal L_{q\bar q}(M^2)}{M^2}\,,
\end{equation}
where $D_R=3$, $C_R=4/3$ and $Q=2/3$. The parton luminosity for a pair of partons $a$ and $b$ is defined as
\begin{equation}
\mathcal L_{ab}(\hat s) = \frac{\hat s}{s}\,\int_{\hat s/s}^1 \frac{dx}{x}\, f_{a/p}(x)\, f_{b/p}\left(\frac{\hat s}{xs}\right)\,,
\end{equation}
where $f_{a,b/p}$ are the relevant PDFs and $\sqrt s$ is the collider centre-of-mass energy.

\item \textbf{Production in association with a gluon.} The cross-section for production in association with a gluon is
\begin{equation}
\sigma (gg \to \Psi^1 g) = \frac{5\pi}{192\,m^2}\,\frac{C_R^3}{D_R}\,\alpha_s^3\bar\alpha_s^3 \int_0^1 dx_1 \int_0^1 dx_2\; f_{g/p}(x_1)\, f_{g/p}(x_2)\; I\left(\frac{x_1 x_2 s}{M^2}\right)\,,
\end{equation}
where 
\begin{equation}
I(x) = \theta(x-1)\left[\frac{2}{x^2}\left(\frac{x+1}{x-1} - \frac{2x\ln x}{(x-1)^2}\right) + \frac{2(x-1)}{x(x+1)^2} + \frac{4\ln x}{(x+1)^3}\right]\,.
\end{equation}
Production in association with a photon or a $Z$ boson is subleading and not considered here. 
\end{enumerate}
For $m_{\psi} \approx 500\GeV$ and $\sqrt{s}=13\TeV$, the electroweak production constitutes almost $80\%$ of the total cross-section, due to the running of the couplings and the phase-space suppression of $gg \to \Psi^1 g$.

To determine the cross-section \cref{eq:partonic_cs_psionium}, we need to compute the branching fraction of the $\Psi^1$ bound state into the following states: 
\begin{enumerate}[i)]
	\item \textbf{SM fermions.} The decay can proceed through a photon or a $Z$ boson and the rate for fermions $f_L, f_R$ is given by
\begin{equation}
\Gamma_{\Psi^1 \to f\bar f} = \frac{n_c}{12}\,D_R C_R^3\sum_{\sigma=R,L}\left(\frac{Y_{f_\sigma} Y}{\cos^2\theta_W}\right)^2
\alpha^2\bar\alpha_s^3 m\,,
\label{eq:annih-ffbar}
\end{equation}
where $n_c = 1$ for leptons and $3$ for quarks.
\item \textbf{$ggg$ or $\gamma gg$.} The decay rates to three gauge bosons (which leads to the bulk of the hadronic decay modes) are
\begin{align}
\Gamma_{\Psi^1 \to ggg} &= \frac{5\left(\pi^2-9\right)}{27\pi}\,\frac{C_R^3}{D_R}\,\alpha_s^3\bar\alpha_s^3 m\,, \\
\Gamma_{\Psi^1 \to \gamma gg} &= \frac{\left(\pi^2-9\right)}{12\pi}\,C_R^5 D_R Q^2\,\alpha \alpha_s^2\bar\alpha_s^3 m\,.
\end{align}
\end{enumerate}
Assuming that there are no other decays with considerable rates, the branching ratio to any single flavor of leptons is around $10\%$. 

\end{appendices}

\bibliographystyle{JHEP}
	
\bibliography{main}

\providecommand{\href}[2]{#2}\begingroup\raggedright\begin{thebibliography}{10}

\bibitem{Bertone:2004pz}
G.~Bertone, D.~Hooper, and J.~Silk, {\it {Particle dark matter: Evidence,
  candidates and constraints}},  {\em Phys. Rept.} {\bf 405} (2005) 279--390,
  [\href{http://arxiv.org/abs/hep-ph/0404175}{{\tt hep-ph/0404175}}].

\bibitem{Hiller:2014yaa}
G.~Hiller and M.~Schmaltz, {\it {$R_K$ and future $b \to s \ell \ell$ physics
  beyond the standard model opportunities}},  {\em Phys. Rev. D} {\bf 90}
  (2014) 054014, [\href{http://arxiv.org/abs/1408.1627}{{\tt
  arXiv:1408.1627}}].

\bibitem{Gripaios:2014tna}
B.~Gripaios, M.~Nardecchia, and S.~A. Renner, {\it {Composite leptoquarks and
  anomalies in $B$-meson decays}},  {\em JHEP} {\bf 05} (2015) 006,
  [\href{http://arxiv.org/abs/1412.1791}{{\tt arXiv:1412.1791}}].

\bibitem{Alonso:2015sja}
R.~Alonso, B.~Grinstein, and J.~Martin~Camalich, {\it {Lepton universality
  violation and lepton flavor conservation in $B$-meson decays}},  {\em JHEP}
  {\bf 10} (2015) 184, [\href{http://arxiv.org/abs/1505.05164}{{\tt
  arXiv:1505.05164}}].

\bibitem{Fajfer:2015ycq}
S.~Fajfer and N.~Ko\v{s}nik, {\it {Vector leptoquark resolution of $R_K$ and
  $R_{D^{(*)}}$ puzzles}},  {\em Phys. Lett. B} {\bf 755} (2016) 270--274,
  [\href{http://arxiv.org/abs/1511.06024}{{\tt arXiv:1511.06024}}].

\bibitem{Calibbi:2015kma}
L.~Calibbi, A.~Crivellin, and T.~Ota, {\it {Effective Field Theory Approach to
  $b\to s\ell\ell^{(')}$, $B\to K^{(*)}\nu\overline{\nu}$ and $B\to
  D^{(*)}\tau\nu$ with Third Generation Couplings}},  {\em Phys. Rev. Lett.}
  {\bf 115} (2015) 181801, [\href{http://arxiv.org/abs/1506.02661}{{\tt
  arXiv:1506.02661}}].

\bibitem{Bauer:2015knc}
M.~Bauer and M.~Neubert, {\it {Minimal Leptoquark Explanation for the
  $R_{D^{(*)}}$ , $R_K$ , and $(g-2)_\mu$ Anomalies}},  {\em Phys. Rev. Lett.}
  {\bf 116} (2016), no.~14 141802, [\href{http://arxiv.org/abs/1511.01900}{{\tt
  arXiv:1511.01900}}].

\bibitem{Barbieri:2015yvd}
R.~Barbieri, G.~Isidori, A.~Pattori, and F.~Senia, {\it {Anomalies in
  $B$-decays and $U(2)$ flavour symmetry}},  {\em Eur. Phys. J. C} {\bf 76}
  (2016), no.~2 67, [\href{http://arxiv.org/abs/1512.01560}{{\tt
  arXiv:1512.01560}}].

\bibitem{Faroughy:2016osc}
D.~A. Faroughy, A.~Greljo, and J.~F. Kamenik, {\it {Confronting lepton flavor
  universality violation in B decays with high-$p_T$ tau lepton searches at
  LHC}},  {\em Phys. Lett. B} {\bf 764} (2017) 126--134,
  [\href{http://arxiv.org/abs/1609.07138}{{\tt arXiv:1609.07138}}].

\bibitem{Becirevic:2016yqi}
D.~Be\v{c}irevi\'c, S.~Fajfer, N.~Ko\v{s}nik, and O.~Sumensari, {\it
  {Leptoquark model to explain the $B$-physics anomalies, $R_K$ and $R_D$}},
  {\em Phys. Rev. D} {\bf 94} (2016), no.~11 115021,
  [\href{http://arxiv.org/abs/1608.08501}{{\tt arXiv:1608.08501}}].

\bibitem{Bhattacharya:2016mcc}
B.~Bhattacharya, A.~Datta, J.-P. Gu\'evin, D.~London, and R.~Watanabe, {\it
  {Simultaneous Explanation of the $R_K$ and $R_{D^{(*)}}$ Puzzles: a Model
  Analysis}},  {\em JHEP} {\bf 01} (2017) 015,
  [\href{http://arxiv.org/abs/1609.09078}{{\tt arXiv:1609.09078}}].

\bibitem{Assad:2017iib}
N.~Assad, B.~Fornal, and B.~Grinstein, {\it {Baryon Number and Lepton
  Universality Violation in Leptoquark and Diquark Models}},  {\em Phys. Lett.
  B} {\bf 777} (2018) 324--331, [\href{http://arxiv.org/abs/1708.06350}{{\tt
  arXiv:1708.06350}}].

\bibitem{Calibbi:2017qbu}
L.~Calibbi, A.~Crivellin, and T.~Li, {\it {Model of vector leptoquarks in view
  of the $B$-physics anomalies}},  {\em Phys. Rev. D} {\bf 98} (2018), no.~11
  115002, [\href{http://arxiv.org/abs/1709.00692}{{\tt arXiv:1709.00692}}].

\bibitem{Blanke:2018sro}
M.~Blanke and A.~Crivellin, {\it {$B$ Meson Anomalies in a Pati-Salam Model
  within the Randall-Sundrum Background}},  {\em Phys. Rev. Lett.} {\bf 121}
  (2018), no.~1 011801, [\href{http://arxiv.org/abs/1801.07256}{{\tt
  arXiv:1801.07256}}].

\bibitem{Becirevic:2018afm}
D.~Be\v{c}irevi\'c, I.~Dor\v{s}ner, S.~Fajfer, N.~Ko\v{s}nik, D.~A. Faroughy,
  and O.~Sumensari, {\it {Scalar leptoquarks from grand unified theories to
  accommodate the $B$-physics anomalies}},  {\em Phys. Rev. D} {\bf 98} (2018),
  no.~5 055003, [\href{http://arxiv.org/abs/1806.05689}{{\tt
  arXiv:1806.05689}}].

\bibitem{Kumar:2018kmr}
J.~Kumar, D.~London, and R.~Watanabe, {\it {Combined Explanations of the $b \to
  s \mu^+ \mu^-$ and $b \to c \tau^- {\bar\nu}$ Anomalies: a General Model
  Analysis}},  {\em Phys. Rev. D} {\bf 99} (2019), no.~1 015007,
  [\href{http://arxiv.org/abs/1806.07403}{{\tt arXiv:1806.07403}}].

\bibitem{Buttazzo:2017ixm}
D.~Buttazzo, A.~Greljo, G.~Isidori, and D.~Marzocca, {\it {B-physics anomalies:
  a guide to combined explanations}},  {\em JHEP} {\bf 11} (2017) 044,
  [\href{http://arxiv.org/abs/1706.07808}{{\tt arXiv:1706.07808}}].

\bibitem{Angelescu:2018tyl}
A.~Angelescu, D.~Be\v{c}irevi\'c, D.~A. Faroughy, and O.~Sumensari, {\it
  {Closing the window on single leptoquark solutions to the $B$-physics
  anomalies}},  {\em JHEP} {\bf 10} (2018) 183,
  [\href{http://arxiv.org/abs/1808.08179}{{\tt arXiv:1808.08179}}].

\bibitem{Angelescu:2021lln}
A.~Angelescu, D.~Be\v{c}irevi\'c, D.~A. Faroughy, F.~Jaffredo, and
  O.~Sumensari, {\it {Single leptoquark solutions to the B-physics anomalies}},
   {\em Phys. Rev. D} {\bf 104} (2021), no.~5 055017,
  [\href{http://arxiv.org/abs/2103.12504}{{\tt arXiv:2103.12504}}].

\bibitem{DiLuzio:2017vat}
L.~Di~Luzio, A.~Greljo, and M.~Nardecchia, {\it {Gauge leptoquark as the origin
  of B-physics anomalies}},  {\em Phys. Rev. D} {\bf 96} (2017), no.~11 115011,
  [\href{http://arxiv.org/abs/1708.08450}{{\tt arXiv:1708.08450}}].

\bibitem{Bordone:2017bld}
M.~Bordone, C.~Cornella, J.~Fuentes-Martin, and G.~Isidori, {\it {A three-site
  gauge model for flavor hierarchies and flavor anomalies}},  {\em Phys. Lett.
  B} {\bf 779} (2018) 317--323, [\href{http://arxiv.org/abs/1712.01368}{{\tt
  arXiv:1712.01368}}].

\bibitem{Greljo:2018tuh}
A.~Greljo and B.~A. Stefanek, {\it {Third family quark\textendash{}lepton
  unification at the TeV scale}},  {\em Phys. Lett. B} {\bf 782} (2018)
  131--138, [\href{http://arxiv.org/abs/1802.04274}{{\tt arXiv:1802.04274}}].

\bibitem{Bordone:2018nbg}
M.~Bordone, C.~Cornella, J.~Fuentes-Mart\'\i{}n, and G.~Isidori, {\it
  {Low-energy signatures of the $\mathrm{PS}^3$ model: from $B$-physics
  anomalies to LFV}},  {\em JHEP} {\bf 10} (2018) 148,
  [\href{http://arxiv.org/abs/1805.09328}{{\tt arXiv:1805.09328}}].

\bibitem{DiLuzio:2018zxy}
L.~Di~Luzio, J.~Fuentes-Martin, A.~Greljo, M.~Nardecchia, and S.~Renner, {\it
  {Maximal Flavour Violation: a Cabibbo mechanism for leptoquarks}},  {\em
  JHEP} {\bf 11} (2018) 081, [\href{http://arxiv.org/abs/1808.00942}{{\tt
  arXiv:1808.00942}}].

\bibitem{Cornella:2019hct}
C.~Cornella, J.~Fuentes-Martin, and G.~Isidori, {\it {Revisiting the vector
  leptoquark explanation of the B-physics anomalies}},  {\em JHEP} {\bf 07}
  (2019) 168, [\href{http://arxiv.org/abs/1903.11517}{{\tt arXiv:1903.11517}}].

\bibitem{Fuentes-Martin:2020bnh}
J.~Fuentes-Mart\'\i{}n and P.~Stangl, {\it {Third-family quark-lepton
  unification with a fundamental composite Higgs}},  {\em Phys. Lett. B} {\bf
  811} (2020) 135953, [\href{http://arxiv.org/abs/2004.11376}{{\tt
  arXiv:2004.11376}}].

\bibitem{King:2021jeo}
S.~F. King, {\it {Twin Pati-Salam theory of flavour with a TeV scale vector
  leptoquark}},  \href{http://arxiv.org/abs/2106.03876}{{\tt
  arXiv:2106.03876}}.

\bibitem{Baker:2019sli}
M.~J. Baker, J.~Fuentes-Mart\'\i{}n, G.~Isidori, and M.~K\"onig, {\it {High-
  $p_T$ signatures in vector\textendash{}leptoquark models}},  {\em Eur. Phys.
  J. C} {\bf 79} (2019), no.~4 334,
  [\href{http://arxiv.org/abs/1901.10480}{{\tt arXiv:1901.10480}}].

\bibitem{Cornella:2021sby}
C.~Cornella, D.~A. Faroughy, J.~Fuentes-Martin, G.~Isidori, and M.~Neubert,
  {\it {Reading the footprints of the B-meson flavor anomalies}},
  \href{http://arxiv.org/abs/2103.16558}{{\tt arXiv:2103.16558}}.

\bibitem{Sierra:2015fma}
D.~Aristizabal~Sierra, F.~Staub, and A.~Vicente, {\it {Shedding light on the
  $b\to s$ anomalies with a dark sector}},  {\em Phys. Rev. D} {\bf 92} (2015),
  no.~1 015001, [\href{http://arxiv.org/abs/1503.06077}{{\tt
  arXiv:1503.06077}}].

\bibitem{Belanger:2015nma}
G.~B\'elanger, C.~Delaunay, and S.~Westhoff, {\it {A Dark Matter Relic From
  Muon Anomalies}},  {\em Phys. Rev. D} {\bf 92} (2015) 055021,
  [\href{http://arxiv.org/abs/1507.06660}{{\tt arXiv:1507.06660}}].

\bibitem{Kawamura:2017ecz}
J.~Kawamura, S.~Okawa, and Y.~Omura, {\it {Interplay between the b$\to
  s\ell\ell$ anomalies and dark matter physics}},  {\em Phys. Rev. D} {\bf 96}
  (2017), no.~7 075041, [\href{http://arxiv.org/abs/1706.04344}{{\tt
  arXiv:1706.04344}}].

\bibitem{Ko:2017quv}
P.~Ko, T.~Nomura, and H.~Okada, {\it {A flavor dependent gauge symmetry,
  Predictive radiative seesaw and LHCb anomalies}},  {\em Phys. Lett. B} {\bf
  772} (2017) 547--552, [\href{http://arxiv.org/abs/1701.05788}{{\tt
  arXiv:1701.05788}}].

\bibitem{Fuyuto:2017sys}
K.~Fuyuto, H.-L. Li, and J.-H. Yu, {\it {Implications of hidden gauged $U(1)$
  model for $B$ anomalies}},  {\em Phys. Rev. D} {\bf 97} (2018), no.~11
  115003, [\href{http://arxiv.org/abs/1712.06736}{{\tt arXiv:1712.06736}}].

\bibitem{Cline:2017qqu}
J.~M. Cline and J.~M. Cornell, {\it {$R({K^{(*)}})$ from dark matter
  exchange}},  {\em Phys. Lett. B} {\bf 782} (2018) 232--237,
  [\href{http://arxiv.org/abs/1711.10770}{{\tt arXiv:1711.10770}}].

\bibitem{Azatov:2018kzb}
A.~Azatov, D.~Barducci, D.~Ghosh, D.~Marzocca, and L.~Ubaldi, {\it {Combined
  explanations of B-physics anomalies: the sterile neutrino solution}},  {\em
  JHEP} {\bf 10} (2018) 092, [\href{http://arxiv.org/abs/1807.10745}{{\tt
  arXiv:1807.10745}}].

\bibitem{Choi:2018stw}
S.-M. Choi, Y.-J. Kang, H.~M. Lee, and T.-G. Ro, {\it {Lepto-Quark Portal Dark
  Matter}},  {\em JHEP} {\bf 10} (2018) 104,
  [\href{http://arxiv.org/abs/1807.06547}{{\tt arXiv:1807.06547}}].

\bibitem{Singirala:2018mio}
S.~Singirala, S.~Sahoo, and R.~Mohanta, {\it {Exploring dark matter, neutrino
  mass and $R_{K^{(*)},\phi}$ anomalies in $L_{\mu}-L_{\tau}$ model}},  {\em
  Phys. Rev. D} {\bf 99} (2019), no.~3 035042,
  [\href{http://arxiv.org/abs/1809.03213}{{\tt arXiv:1809.03213}}].

\bibitem{Hati:2018fzc}
C.~Hati, G.~Kumar, J.~Orloff, and A.~M. Teixeira, {\it {Reconciling $B$-meson
  decay anomalies with neutrino masses, dark matter and constraints from
  flavour violation}},  {\em JHEP} {\bf 11} (2018) 011,
  [\href{http://arxiv.org/abs/1806.10146}{{\tt arXiv:1806.10146}}].

\bibitem{Falkowski:2018dsl}
A.~Falkowski, S.~F. King, E.~Perdomo, and M.~Pierre, {\it {Flavourful $Z'$
  portal for vector-like neutrino Dark Matter and $R_{K^{(*)}}$}},  {\em JHEP}
  {\bf 08} (2018) 061, [\href{http://arxiv.org/abs/1803.04430}{{\tt
  arXiv:1803.04430}}].

\bibitem{Baek:2018aru}
S.~Baek and C.~Yu, {\it {Dark matter for $b\to s \mu^+ \mu^-$ anomaly in a
  gauged $U(1)_X$ model}},  {\em JHEP} {\bf 11} (2018) 054,
  [\href{http://arxiv.org/abs/1806.05967}{{\tt arXiv:1806.05967}}].

\bibitem{Hutauruk:2019crc}
P.~T.~P. Hutauruk, T.~Nomura, H.~Okada, and Y.~Orikasa, {\it {Dark matter and
  $B$-meson anomalies in a flavor dependent gauge symmetry}},  {\em Phys. Rev.
  D} {\bf 99} (2019), no.~5 055041,
  [\href{http://arxiv.org/abs/1901.03932}{{\tt arXiv:1901.03932}}].

\bibitem{Trifinopoulos:2019lyo}
S.~Trifinopoulos, {\it {B -physics anomalies: The bridge between R -parity
  violating supersymmetry and flavored dark matter}},  {\em Phys. Rev. D} {\bf
  100} (2019), no.~11 115022, [\href{http://arxiv.org/abs/1904.12940}{{\tt
  arXiv:1904.12940}}].

\bibitem{Guadagnoli:2020tlx}
D.~Guadagnoli, M.~Reboud, and P.~Stangl, {\it {The Dark Side of 4321}},  {\em
  JHEP} {\bf 10} (2020) 084, [\href{http://arxiv.org/abs/2005.10117}{{\tt
  arXiv:2005.10117}}].

\bibitem{Carvunis:2020exc}
A.~Carvunis, D.~Guadagnoli, M.~Reboud, and P.~Stangl, {\it {Composite Dark
  Matter and a horizontal symmetry}},  {\em JHEP} {\bf 02} (2021) 056,
  [\href{http://arxiv.org/abs/2007.11931}{{\tt arXiv:2007.11931}}].

\bibitem{Huang:2020ris}
D.~Huang, A.~P. Morais, and R.~Santos, {\it {Anomalies in $B$-meson decays and
  the muon $g-2$ from dark loops}},  {\em Phys. Rev. D} {\bf 102} (2020), no.~7
  075009, [\href{http://arxiv.org/abs/2007.05082}{{\tt arXiv:2007.05082}}].

\bibitem{DEramo:2020sqv}
F.~D'Eramo, N.~Ko\v{s}nik, F.~Pobbe, A.~Smolkovi\v{c}, and O.~Sumensari, {\it
  {Leptoquarks and real singlets: A richer scalar sector behind the origin of
  dark matter}},  {\em Phys. Rev. D} {\bf 104} (2021), no.~1 015035,
  [\href{http://arxiv.org/abs/2012.05743}{{\tt arXiv:2012.05743}}].

\bibitem{Arcadi:2021glq}
G.~Arcadi, L.~Calibbi, M.~Fedele, and F.~Mescia, {\it {Systematic approach to
  $B$-physics anomalies and $t$-channel dark matter}},
  \href{http://arxiv.org/abs/2103.09835}{{\tt arXiv:2103.09835}}.

\bibitem{Becker:2021sfd}
M.~Becker, D.~D\"oring, S.~Karmakar, and H.~P\"as, {\it {Fermionic Singlet Dark
  Matter in One-Loop Solutions to the $R_K$ Anomaly: A Systematic Study}},
  \href{http://arxiv.org/abs/2103.12043}{{\tt arXiv:2103.12043}}.

\bibitem{Arcadi:2021cwg}
G.~Arcadi, L.~Calibbi, M.~Fedele, and F.~Mescia, {\it {Muon $g-2$ and
  $B$-anomalies from Dark Matter}},  {\em Phys. Rev. Lett.} {\bf 127} (2021),
  no.~6 061802, [\href{http://arxiv.org/abs/2104.03228}{{\tt
  arXiv:2104.03228}}].

\bibitem{Pati:1974yy}
J.~C. Pati and A.~Salam, {\it {Lepton Number as the Fourth Color}},  {\em Phys.
  Rev. D} {\bf 10} (1974) 275--289. [Erratum: Phys.Rev.D 11, 703--703 (1975)].

\bibitem{DeSimone:2010tf}
A.~De~Simone, V.~Sanz, and H.~P. Sato, {\it {Pseudo-Dirac Dark Matter Leaves a
  Trace}},  {\em Phys. Rev. Lett.} {\bf 105} (2010) 121802,
  [\href{http://arxiv.org/abs/1004.1567}{{\tt arXiv:1004.1567}}].

\bibitem{ATLAS:2018uni}
{\bf ATLAS} Collaboration, M.~Aaboud et~al., {\it {Search for resonant and
  non-resonant Higgs boson pair production in the ${b\bar{b}\tau^+\tau^-}$
  decay channel in $pp$ collisions at $\sqrt{s}=13$ TeV with the ATLAS
  detector}},  {\em Phys. Rev. Lett.} {\bf 121} (2018), no.~19 191801,
  [\href{http://arxiv.org/abs/1808.00336}{{\tt arXiv:1808.00336}}]. [Erratum:
  Phys.Rev.Lett. 122, 089901 (2019)].

\bibitem{ATLAS:2019qpq}
{\bf ATLAS} Collaboration, M.~Aaboud et~al., {\it {Searches for
  third-generation scalar leptoquarks in $\sqrt{s}$ = 13 TeV pp collisions with
  the ATLAS detector}},  {\em JHEP} {\bf 06} (2019) 144,
  [\href{http://arxiv.org/abs/1902.08103}{{\tt arXiv:1902.08103}}].

\bibitem{CMS:2018txo}
{\bf CMS} Collaboration, A.~M. Sirunyan et~al., {\it {Search for a singly
  produced third-generation scalar leptoquark decaying to a $\tau$ lepton and a
  bottom quark in proton-proton collisions at $\sqrt{s} =$ 13 TeV}},  {\em
  JHEP} {\bf 07} (2018) 115, [\href{http://arxiv.org/abs/1806.03472}{{\tt
  arXiv:1806.03472}}].

\bibitem{CMS:2018iye}
{\bf CMS} Collaboration, A.~M. Sirunyan et~al., {\it {Search for heavy
  neutrinos and third-generation leptoquarks in hadronic states of two $\tau$
  leptons and two jets in proton-proton collisions at $\sqrt{s} =$ 13 TeV}},
  {\em JHEP} {\bf 03} (2019) 170, [\href{http://arxiv.org/abs/1811.00806}{{\tt
  arXiv:1811.00806}}].

\bibitem{Tucker-Smith:2001myb}
D.~Tucker-Smith and N.~Weiner, {\it {Inelastic dark matter}},  {\em Phys. Rev.
  D} {\bf 64} (2001) 043502, [\href{http://arxiv.org/abs/hep-ph/0101138}{{\tt
  hep-ph/0101138}}].

\bibitem{Baker:2015qna}
M.~J. Baker et~al., {\it {The Coannihilation Codex}},  {\em JHEP} {\bf 12}
  (2015) 120, [\href{http://arxiv.org/abs/1510.03434}{{\tt arXiv:1510.03434}}].

\bibitem{Griest:1990kh}
K.~Griest and D.~Seckel, {\it {Three exceptions in the calculation of relic
  abundances}},  {\em Phys. Rev. D} {\bf 43} (1991) 3191--3203.

\bibitem{Alloul:2013bka}
A.~Alloul, N.~D. Christensen, C.~Degrande, C.~Duhr, and B.~Fuks, {\it
  {FeynRules 2.0 - A complete toolbox for tree-level phenomenology}},  {\em
  Comput. Phys. Commun.} {\bf 185} (2014) 2250--2300,
  [\href{http://arxiv.org/abs/1310.1921}{{\tt arXiv:1310.1921}}].

\bibitem{Belanger:2018mqt}
G.~B\'elanger, F.~Boudjema, A.~Goudelis, A.~Pukhov, and B.~Zaldivar, {\it
  {micrOMEGAs5.0 : Freeze-in}},  {\em Comput. Phys. Commun.} {\bf 231} (2018)
  173--186, [\href{http://arxiv.org/abs/1801.03509}{{\tt arXiv:1801.03509}}].

\bibitem{Buschmann:2016hkc}
M.~Buschmann, S.~El~Hedri, A.~Kaminska, J.~Liu, M.~de~Vries, X.-P. Wang, F.~Yu,
  and J.~Zurita, {\it {Hunting for dark matter coannihilation by mixing dijet
  resonances and missing transverse energy}},  {\em JHEP} {\bf 09} (2016) 033,
  [\href{http://arxiv.org/abs/1605.08056}{{\tt arXiv:1605.08056}}].

\bibitem{Keung:2017kot}
W.-Y. Keung, I.~Low, and Y.~Zhang, {\it {Reappraisal of dark matter
  co-annihilating with a top or bottom partner}},  {\em Phys. Rev. D} {\bf 96}
  (2017), no.~1 015008, [\href{http://arxiv.org/abs/1703.02977}{{\tt
  arXiv:1703.02977}}].

\bibitem{ElHedri:2017nny}
S.~El~Hedri, A.~Kaminska, M.~de~Vries, and J.~Zurita, {\it {Simplified
  Phenomenology for Colored Dark Sectors}},  {\em JHEP} {\bf 04} (2017) 118,
  [\href{http://arxiv.org/abs/1703.00452}{{\tt arXiv:1703.00452}}].

\bibitem{Aghanim:2018eyx}
{\bf Planck} Collaboration, N.~Aghanim et~al., {\it {Planck 2018 results. VI.
  Cosmological parameters}},  {\em Astron. Astrophys.} {\bf 641} (2020) A6,
  [\href{http://arxiv.org/abs/1807.06209}{{\tt arXiv:1807.06209}}]. [Erratum:
  Astron.Astrophys. 652, C4 (2021)].

\bibitem{Baker:2018uox}
M.~J. Baker and A.~Thamm, {\it {Leptonic WIMP Coannihilation and the Current
  Dark Matter Search Strategy}},  {\em JHEP} {\bf 10} (2018) 187,
  [\href{http://arxiv.org/abs/1806.07896}{{\tt arXiv:1806.07896}}].

\bibitem{Fabiano:1997xh}
N.~Fabiano, {\it {Top mesons}},  {\em Eur. Phys. J. C} {\bf 2} (1998) 345--350,
  [\href{http://arxiv.org/abs/hep-ph/9704261}{{\tt hep-ph/9704261}}].

\bibitem{Kats:2012ym}
Y.~Kats and M.~J. Strassler, {\it {Probing Colored Particles with Photons,
  Leptons, and Jets}},  {\em JHEP} {\bf 11} (2012) 097,
  [\href{http://arxiv.org/abs/1204.1119}{{\tt arXiv:1204.1119}}]. [Erratum:
  JHEP 07, 009 (2016)].

\bibitem{CMS:2018dqv}
{\bf CMS} Collaboration, A.~M. Sirunyan et~al., {\it {Search for physics beyond
  the standard model in high-mass diphoton events from proton-proton collisions
  at $\sqrt{s} =$ 13 TeV}},  {\em Phys. Rev. D} {\bf 98} (2018), no.~9 092001,
  [\href{http://arxiv.org/abs/1809.00327}{{\tt arXiv:1809.00327}}].

\bibitem{Chetyrkin:2000yt}
K.~G. Chetyrkin, J.~H. Kuhn, and M.~Steinhauser, {\it {RunDec: A Mathematica
  package for running and decoupling of the strong coupling and quark masses}},
   {\em Comput. Phys. Commun.} {\bf 133} (2000) 43--65,
  [\href{http://arxiv.org/abs/hep-ph/0004189}{{\tt hep-ph/0004189}}].

\bibitem{Harland-Lang:2014zoa}
L.~A. Harland-Lang, A.~D. Martin, P.~Motylinski, and R.~S. Thorne, {\it {Parton
  distributions in the LHC era: MMHT 2014 PDFs}},  {\em Eur. Phys. J. C} {\bf
  75} (2015), no.~5 204, [\href{http://arxiv.org/abs/1412.3989}{{\tt
  arXiv:1412.3989}}].

\bibitem{Ball:2014uwa}
{\bf NNPDF} Collaboration, R.~D. Ball et~al., {\it {Parton distributions for
  the LHC Run II}},  {\em JHEP} {\bf 04} (2015) 040,
  [\href{http://arxiv.org/abs/1410.8849}{{\tt arXiv:1410.8849}}].

\bibitem{Butterworth:2015oua}
J.~Butterworth et~al., {\it {PDF4LHC recommendations for LHC Run II}},  {\em J.
  Phys. G} {\bf 43} (2016) 023001, [\href{http://arxiv.org/abs/1510.03865}{{\tt
  arXiv:1510.03865}}].

\bibitem{CMS:2019tbu}
{\bf CMS} Collaboration, {\it {Search for a narrow resonance in high-mass
  dilepton final states in proton-proton collisions using
  140$~\mathrm{fb}^{-1}$ of data at $\sqrt{s}=13~\mathrm{TeV}$}}, .

\bibitem{Chakraborty:2018kqn}
A.~Chakraborty, S.~Kuttimalai, S.~H. Lim, M.~M. Nojiri, and R.~Ruiz, {\it
  {Monojet Signatures from Heavy Colored Particles: Future Collider
  Sensitivities and Theoretical Uncertainties}},  {\em Eur. Phys. J. C} {\bf
  78} (2018), no.~8 679, [\href{http://arxiv.org/abs/1805.05346}{{\tt
  arXiv:1805.05346}}].

\bibitem{ATLAS:2017bfj}
{\bf ATLAS} Collaboration, M.~Aaboud et~al., {\it {Search for dark matter and
  other new phenomena in events with an energetic jet and large missing
  transverse momentum using the ATLAS detector}},  {\em JHEP} {\bf 01} (2018)
  126, [\href{http://arxiv.org/abs/1711.03301}{{\tt arXiv:1711.03301}}].

\bibitem{Degrande:2011ua}
C.~Degrande, C.~Duhr, B.~Fuks, D.~Grellscheid, O.~Mattelaer, and T.~Reiter,
  {\it {UFO - The Universal FeynRules Output}},  {\em Comput. Phys. Commun.}
  {\bf 183} (2012) 1201--1214, [\href{http://arxiv.org/abs/1108.2040}{{\tt
  arXiv:1108.2040}}].

\bibitem{Alwall:2014hca}
J.~Alwall, R.~Frederix, S.~Frixione, V.~Hirschi, F.~Maltoni, O.~Mattelaer,
  H.~S. Shao, T.~Stelzer, P.~Torrielli, and M.~Zaro, {\it {The automated
  computation of tree-level and next-to-leading order differential cross
  sections, and their matching to parton shower simulations}},  {\em JHEP} {\bf
  07} (2014) 079, [\href{http://arxiv.org/abs/1405.0301}{{\tt
  arXiv:1405.0301}}].

\bibitem{Sjostrand:2014zea}
T.~Sj\"ostrand, S.~Ask, J.~R. Christiansen, R.~Corke, N.~Desai, P.~Ilten,
  S.~Mrenna, S.~Prestel, C.~O. Rasmussen, and P.~Z. Skands, {\it {An
  introduction to PYTHIA 8.2}},  {\em Comput. Phys. Commun.} {\bf 191} (2015)
  159--177, [\href{http://arxiv.org/abs/1410.3012}{{\tt arXiv:1410.3012}}].

\bibitem{Cacciari:2011ma}
M.~Cacciari, G.~P. Salam, and G.~Soyez, {\it {FastJet User Manual}},  {\em Eur.
  Phys. J. C} {\bf 72} (2012) 1896, [\href{http://arxiv.org/abs/1111.6097}{{\tt
  arXiv:1111.6097}}].

\bibitem{deFavereau:2013fsa}
{\bf DELPHES 3} Collaboration, J.~de~Favereau, C.~Delaere, P.~Demin,
  A.~Giammanco, V.~Lema\^\i{}tre, A.~Mertens, and M.~Selvaggi, {\it {DELPHES 3,
  A modular framework for fast simulation of a generic collider experiment}},
  {\em JHEP} {\bf 02} (2014) 057, [\href{http://arxiv.org/abs/1307.6346}{{\tt
  arXiv:1307.6346}}].

\bibitem{Larkoski:2013eya}
A.~J. Larkoski, G.~P. Salam, and J.~Thaler, {\it {Energy Correlation Functions
  for Jet Substructure}},  {\em JHEP} {\bf 06} (2013) 108,
  [\href{http://arxiv.org/abs/1305.0007}{{\tt arXiv:1305.0007}}].

\bibitem{Chen2016}
T.~Chen and C.~Guestrin, {\it {XGBoost: {A} Scalable Tree Boosting System}},
  {\em CoRR} (2016) [\href{http://arxiv.org/abs/1603.02754}{{\tt
  arXiv:1603.02754}}].

\bibitem{Read:2002hq}
A.~L. Read, {\it {Presentation of search results: The CL(s) technique}},  {\em
  J. Phys. G} {\bf 28} (2002) 2693--2704.

\bibitem{Cowan:2010js}
G.~Cowan, K.~Cranmer, E.~Gross, and O.~Vitells, {\it {Asymptotic formulae for
  likelihood-based tests of new physics}},  {\em Eur. Phys. J. C} {\bf 71}
  (2011) 1554, [\href{http://arxiv.org/abs/1007.1727}{{\tt arXiv:1007.1727}}].
  [Erratum: Eur.Phys.J.C 73, 2501 (2013)].

\bibitem{pyhf}
L.~Heinrich, M.~Feickert, and G.~Stark, {\it {pyhf: v0.6.2}}, .
  https://github.com/scikit-hep/pyhf/releases/tag/v0.6.2.

\bibitem{pyhfjoss}
L.~Heinrich, M.~Feickert, G.~Stark, and K.~Cranmer, {\it pyhf: pure-python
  implementation of histfactory statistical models},  {\em Journal of Open
  Source Software} {\bf 6} (2021), no.~58 2823.

\bibitem{Fuentes-Martin:2020pww}
J.~Fuentes-Martin, G.~Isidori, J.~Pag\`es, and B.~A. Stefanek, {\it {Flavor
  non-universal Pati-Salam unification and neutrino masses}},  {\em Phys. Lett.
  B} {\bf 820} (2021) 136484, [\href{http://arxiv.org/abs/2012.10492}{{\tt
  arXiv:2012.10492}}].

\bibitem{Allwicher:2020esa}
L.~Allwicher, G.~Isidori, and A.~E. Thomsen, {\it {Stability of the Higgs
  Sector in a Flavor-Inspired Multi-Scale Model}},  {\em JHEP} {\bf 01} (2021)
  191, [\href{http://arxiv.org/abs/2011.01946}{{\tt arXiv:2011.01946}}].

\end{thebibliography}\endgroup
	
\end{document}